\documentclass[twocolumn,pra,aps,showpacs,superscriptaddress]{revtex4}
\usepackage{amsfonts}
\usepackage{amssymb}
\usepackage{amsmath}
\usepackage{graphicx}
\usepackage{amsbsy}
\usepackage[tight]{subfigure}
\begin{document}

\title{Spatio-temporal extreme events in a laser with a saturable absorber}

\author{Cristina Rimoldi}
\author{St\'ephane Barland}
\affiliation{Universit\'e C\^ote d'Azur, CNRS, Institut Non Lin\'eaire de Nice, France\footnote{1361, route des Lucioles, 06560 Valbonne, France}}
\author{Franco Prati}
\affiliation{CNISM, and Dipartimento di Scienza e Alta Tecnologia, Universit\'a dell'Insubria, Via Valleggio 11, 22100 Como, Italy}
\author{Giovanna Tissoni\footnote{Corresponding author: giovanna.tissoni@inln.cnrs.fr}}
\affiliation{Universit\'e C\^ote d'Azur, CNRS, Institut Non Lin\'eaire de Nice, France\footnote{1361, route des Lucioles, 06560 Valbonne, France}}

\date{\today}

\begin{abstract}
We study extreme events occurring in the transverse $(x,y)$ section of the field emitted by a broad-area semiconductor laser with a saturable absorber. The spatio--temporal events on which we perform the statistical analysis are identified as maxima of the field intensity in the 3D space $(x,y,t)$. We identify regions in the parameter space where extreme events are more likely to occur and we study the connection of those extreme events with the cavity solitons that are known to exist in the same system, both stationary and self--pulsing.
\end{abstract}

\maketitle

\section{Introduction}
\label{intro}

In the recent years, extreme events in optics have been attracting a lot of interest, originating from the seminal paper by Solli \textit{et al.} \cite{solli}, due to the well-known analogy between optics and hydrodynamics, where rogue wave formation and prediction is a priority field of investigations.
A huge literature has been blooming in many different optical systems besides optical fibers (for a review, see \cite{onorato,dudley} and references therein).

Very recently, extreme events were studied both experimentally and numerically \cite{selmi} in the intensity emitted by a monolithic broad-area VCSEL with a saturable absorber with a linear pump (which reduces to one the transverse dimensions).
Spatio--temporal chaos is claimed to be at the dynamical origin of extreme events but, nevertheless, no insight about the spatio-temporal nature of these events is given, like typical spatial size or temporal duration.

Here we show numerical results about extreme events occurring in the field intensity emitted by a monolithic broad-area VCSEL with an intra--cavity saturable absorber \cite{bache,prati,moving,VahedEPJD2012,VahedPRSA2014}, as the one used in the experiments on cavity solitons \cite{elsass,selmi}.

We show that below the lasing threshold, the system may present multiple stable solutions, such as stationary cavity solitons, oscillating or chaotic solitons and a global
turbulent solution where the light intensity oscillates aperiodically in space and time, together with the trivial non--lasing solution. The turbulent solution survives above threshold, where it is the only attractor of the system. When the system is emitting on the turbulent state, we perform a statistical treatment on the full set of 3D data of field intensity as a function of space and time.
In contrast with previous literature about optical rogue waves in spatially extended systems \cite{arecchi,oppo,liu,selmi,oppoPRL2016}, we developed a numerical method for the individuation of the spatio-temporal maxima of the transverse field intensity in which each maximum appearing in the space profile is counted as an \textquotedblleft event\textquotedblright only when its peak intensity reaches the maximum value also in time. This method allows a comparison, for example, with the hydrodynamical definition of \textquotedblleft significant wave height\textquotedblright, corresponding to the mean value of the wave height (from trough to crest) of the highest third of the waves.

A comparison with the existing methods of statistical analysis of extreme events in other transverse systems has been also developed.

In conservative systems and propagative geometry, the rogue wave phenomenon has often be related to known solutions of the NLS equation such as Akhmediev breathers or Peregrine solitons \cite{dudley,kedziora2013classifying,kibler2010peregrine}. However, recent measurements in nonlinear optical fiber have indicated that rogue waves may differ significantly from these analytic solutions \cite{randoux2016inverse}. In the present case of a dissipative system, dissipative solitons are attractors of the dynamics and their signature in phase space might be expected to play the role in the formation of rogue waves. For this reason,  we studied the relationship between the spatial size of the rogue waves and that of the stationary solitons and between the temporal behaviour of the rogue waves and that of the oscillating solitons.
We also found correlations among the probability of observing rogue waves and the different stability domains of the solitons.

We believe that our system, being intrinsically two-dimensional, may give some precious insights on the focusing mechanisms giving rise to rogue wave formation in oceans, mechanisms that could be absent in one-dimensional systems such as fibers, where optical rogue waves are mostly studied.

In section \ref{themodel} we recall the dynamical equations that we use to describe a semiconductor laser with an intracavity saturable absorber, while in section \ref{extremeevents} we present our method for the selection of spatio--temporal maxima and define different thresholds for extreme events. In section \ref{parameters} we analyze the dependence of the probability of extreme events on the laser parameters and compare our results with those that would be obtained with the usual method of RW analysis based on total intensity distribution $I(x,y,t)$ \cite{arecchi,oppo,liu,oppoPRL2016}.
Finally, in section \ref{profiles} we analyse the spatial and temporal profiles of the intensity or carrier active/passive populations in the presence of extreme events and we compare them with cavity solitons, both stationary and self--pulsing.

\section{The model}\label{themodel}
We consider a monolithic broad-area VCSEL (Vertical Cavity Surface Emitting Laser) with an intracavity saturable absorber, described by the following set of equations \cite{bache,prati}
\begin{eqnarray}
\dot{F}&=&[(1-i\alpha)D+(1-i\beta)d-1+(\delta+i)\nabla_\bot^2]F\,,\nonumber\\
\dot{D}&=&b[\mu -D(1+|F|^2)-BD^2]\,, \label{model}\\
\dot{d}&=&rb[-\gamma-d(1+s|F|^2)-Bd^2]\,,\nonumber
\end{eqnarray}
where $F$ is the slowly varying amplitude of the electric field, $D$ ($d$) is the population variable related to the carrier density in the active (passive) material; $\mu$ ($\gamma$) is the pump (absorption) parameter, $\alpha$ ($\beta$) is the linewidth enhancement factor of the active (passive) material; $b$ and $r$ are, respectively, the ratio of the photon to the carrier lifetime in the amplifier and the ratio of the carrier lifetimes in the amplifier to the one of the absorber; $B$ is the coefficient of radiative recombination, assumed identical for simplicity in the two materials; $s$ is the saturation parameter, and $\delta$ is a diffusion coefficient for the electric field that accounts phenomenologically for the finite linewidth of gain.

Time is scaled to the photon lifetime ($\approx$ 10 ps) and space is scaled to the diffraction length ($\approx$ 4 $\mu$m). For a more detailed definition of all these parameters see \cite{VahedEPJD2012}.
\begin{figure}
\includegraphics[width=0.75\linewidth]{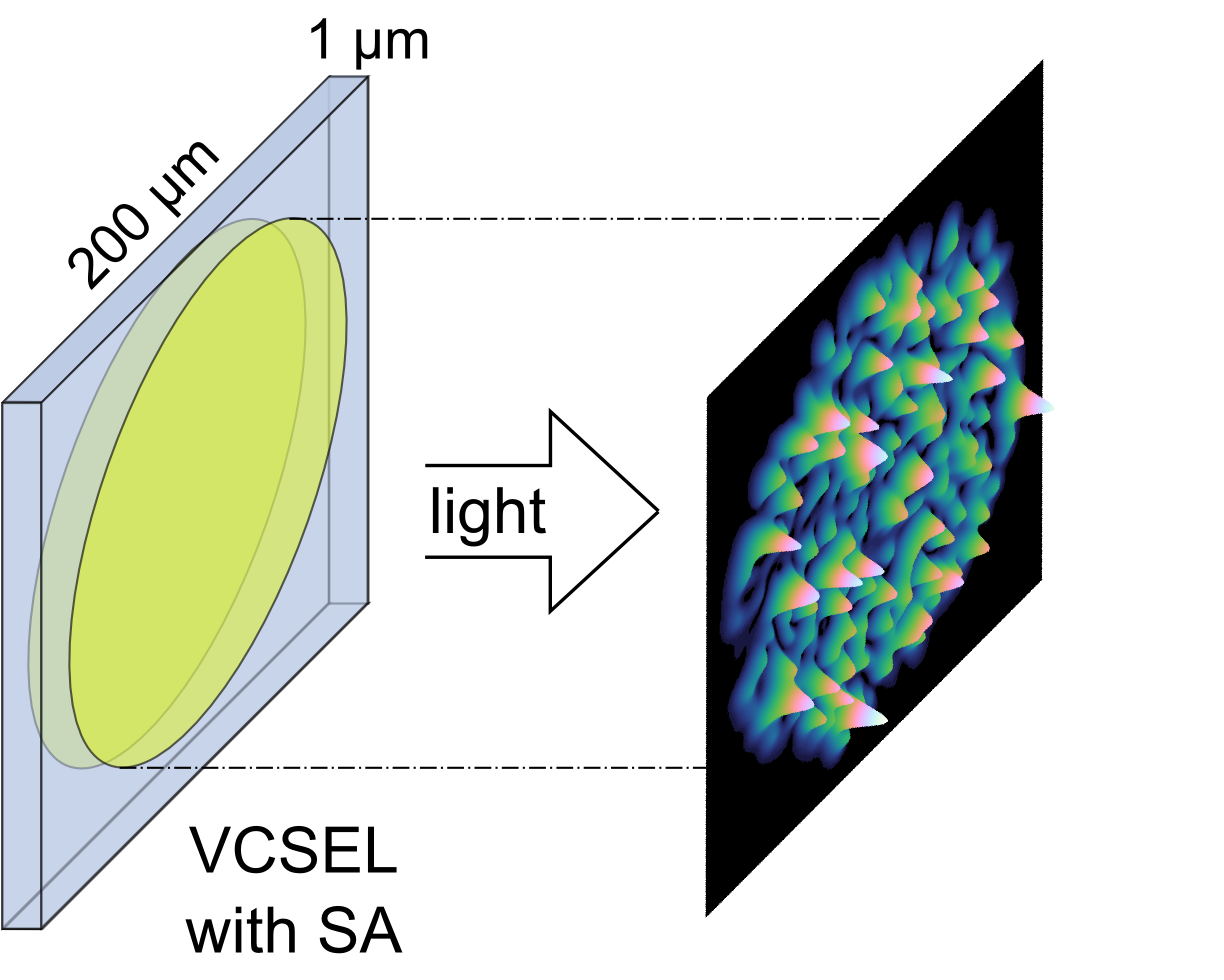}
\caption{A scheme of the system in study: a broad--area VCSEL with an intra--cavity saturable absorber. The spatio--temporal profile of the emitted light is analysed.}
\label{setup}
\end{figure}
\begin{figure}
\centering
\includegraphics[width=0.9\linewidth]{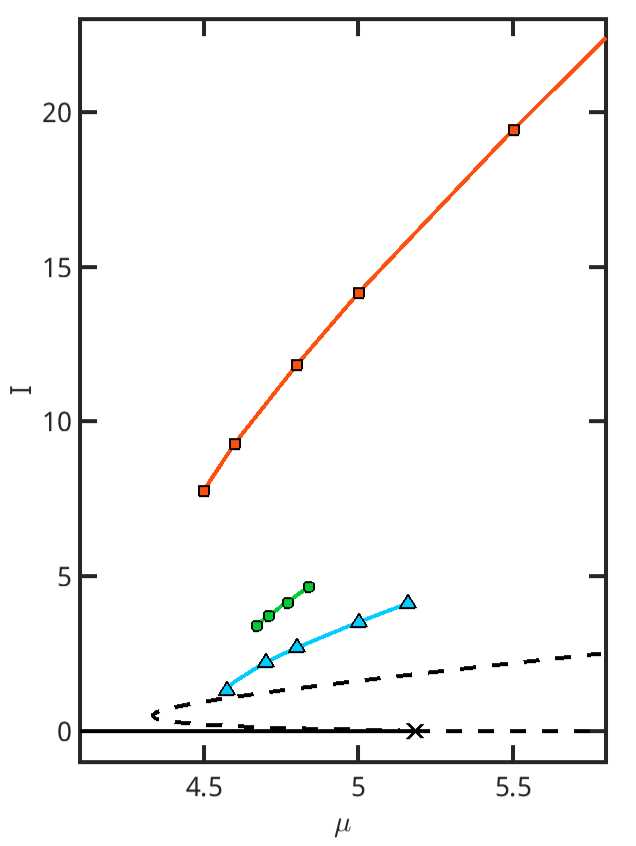}
\caption{Homogenous stationary solution for the system (\ref{model}) (dashed black line), stationary cavity soliton branch (blue line and triangles), time averaged maximum intensity of the turbulent state (orange line and squares) and of chaotic solitons (green line and circles), as a function of $\mu$.
Other parameters are: $r=1$, $b=0.01$, $\alpha=2$, $\beta=1$, $\gamma=2$, $s=1$, $B=0.1$, $\delta=0.01$. The laser threshold is at $\mu_{th} = 5.18$}
\label{stationarystate}
\end{figure}

In \cite{VahedPRSA2014} it was shown that for a large region of the parameter space a spatio-temporal turbulent state coexists below the laser threshold with the non-lasing solution, the stationary cavity solitons and possibly with localised chaotic states (chaotic solitons), Conversely, above the laser threshold, where the non-lasing solution becomes unstable, the extended spatio-temporal turbulent state is the only possible solution of the equations. Throughout all the paper we study the behaviour of the system in such a turbulent state, both below and above the lasing threshold.

\begin{figure}[htb!]
  \centerline{\subfigure[ ]{\includegraphics[width=0.5\columnwidth]{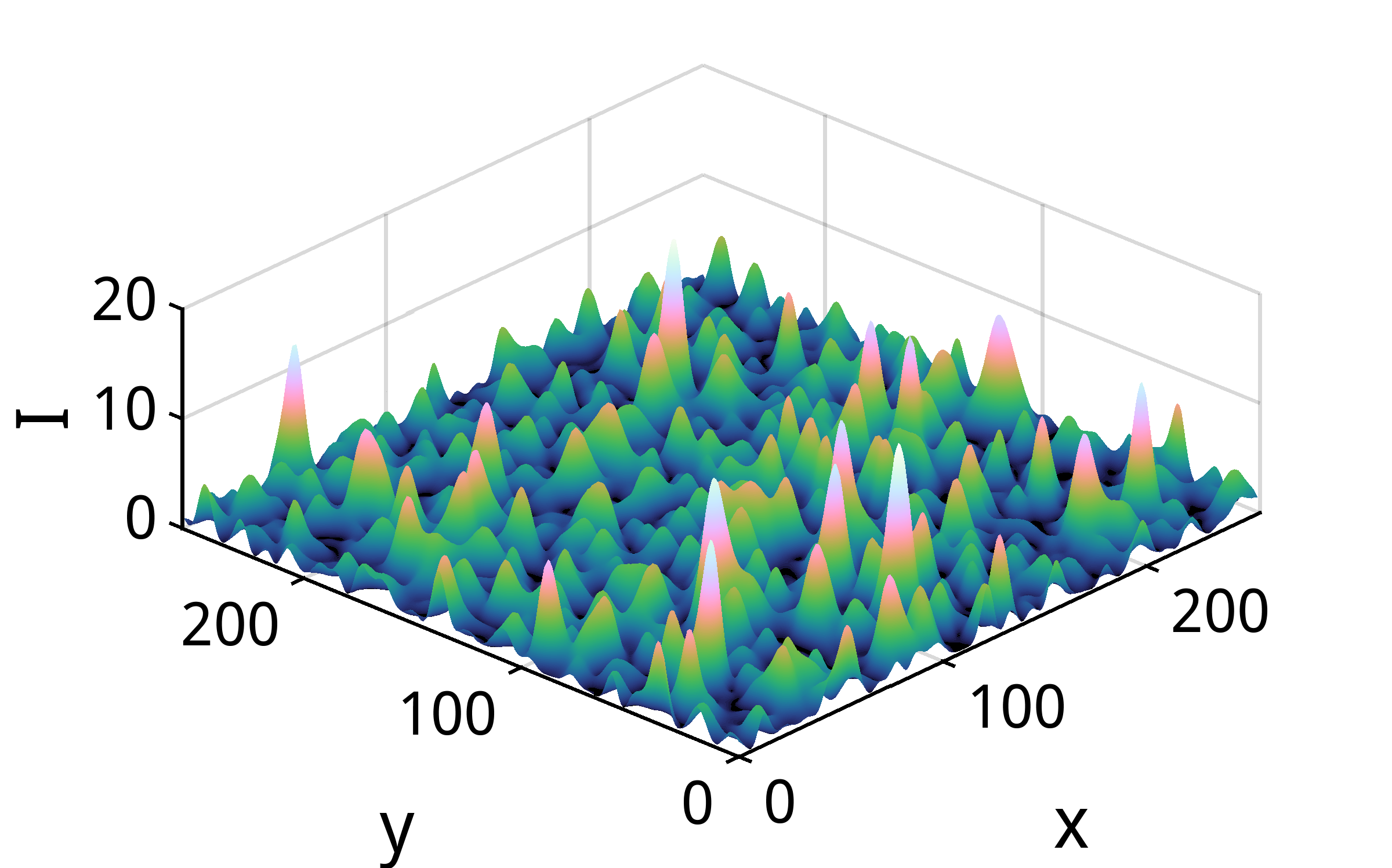}
    }
      \hfil
    \subfigure[ ]{\includegraphics[width=0.5\columnwidth]{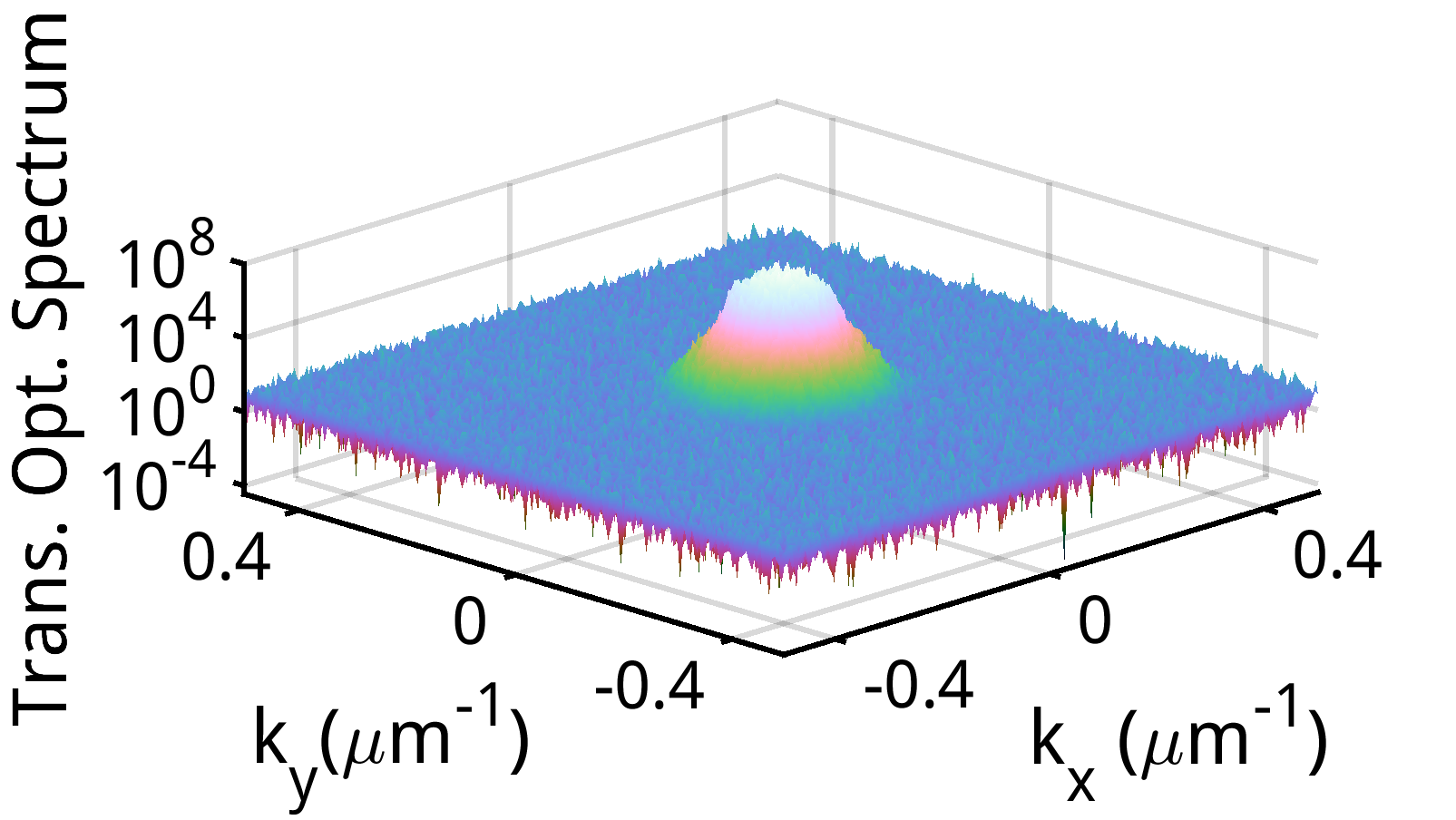}
  }
  }
  \caption{(a) Snapshot of the transverse profile of the field intensity and (b) corresponding Fourier spectrum (in logarithmic scale) for the turbulent solution, for $\mu=5$ and $r=1$.}
  \label{turbulent2}
\end{figure}

\begin{figure}[htb!]
  \centerline{\subfigure[ ]{\includegraphics[width=0.5\columnwidth]{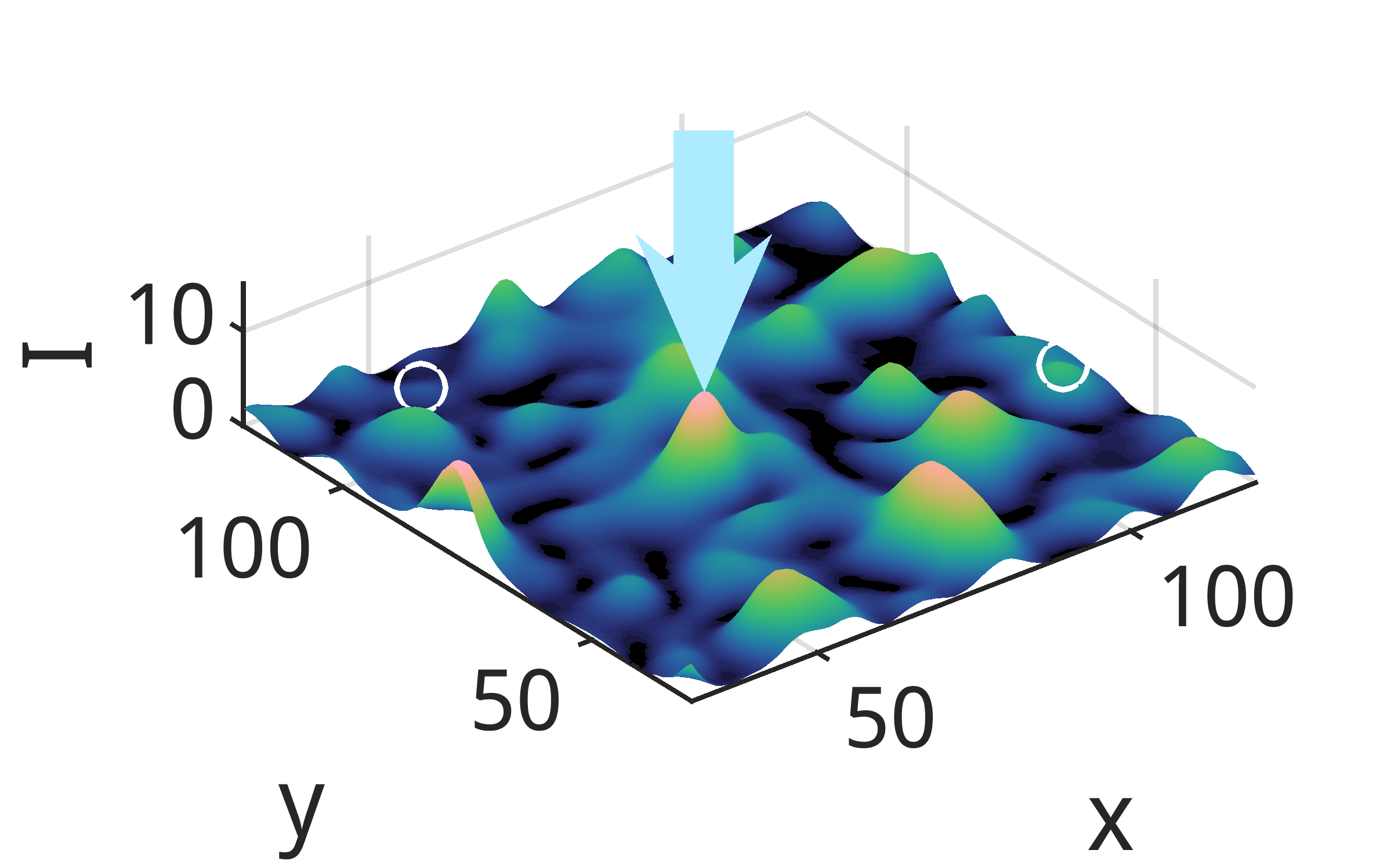}
    }
      \hfil
    \subfigure[ ]{\includegraphics[width=0.5\columnwidth]{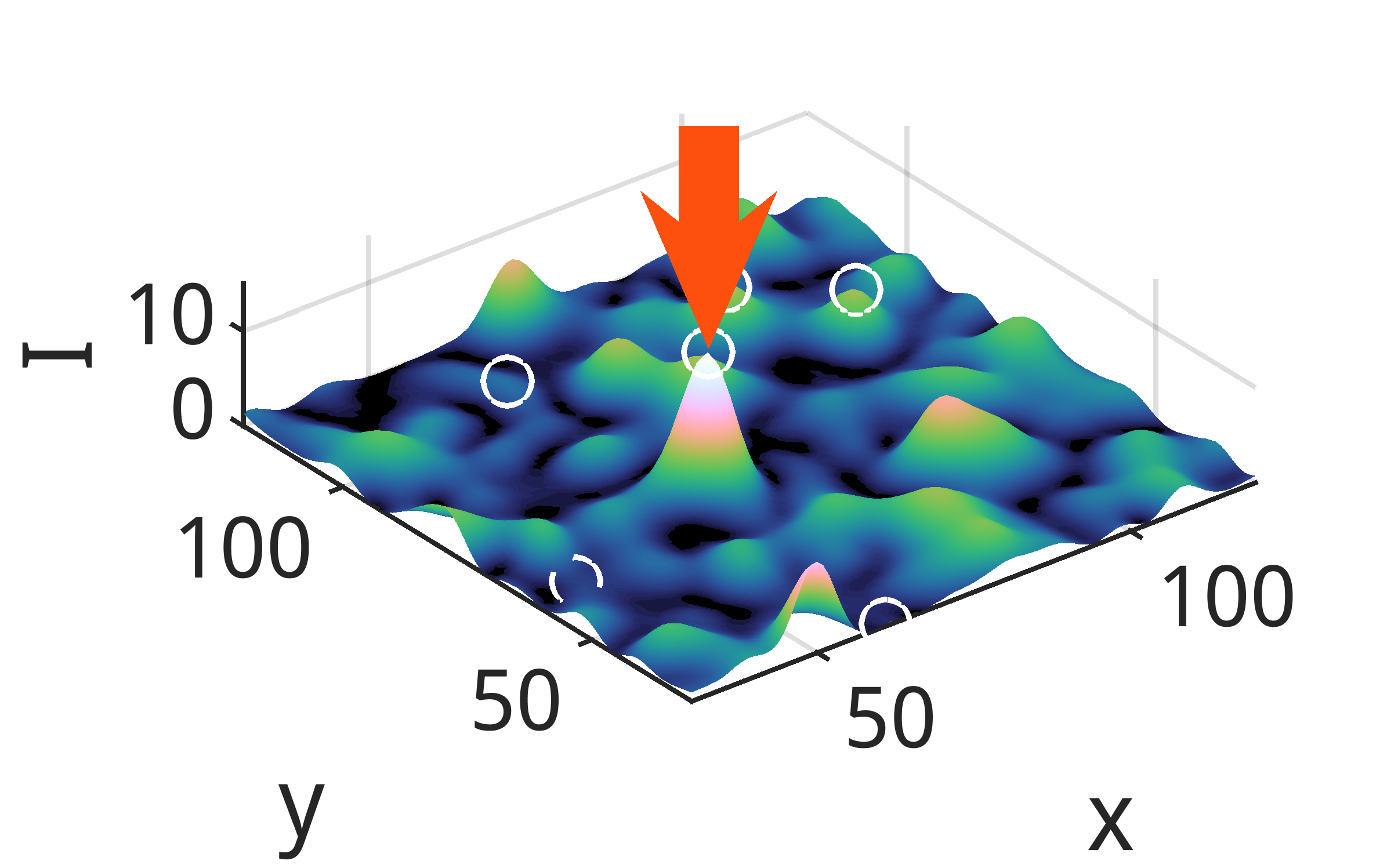}
  }}
    \subfigure[ ]{\includegraphics[width=0.5\columnwidth]{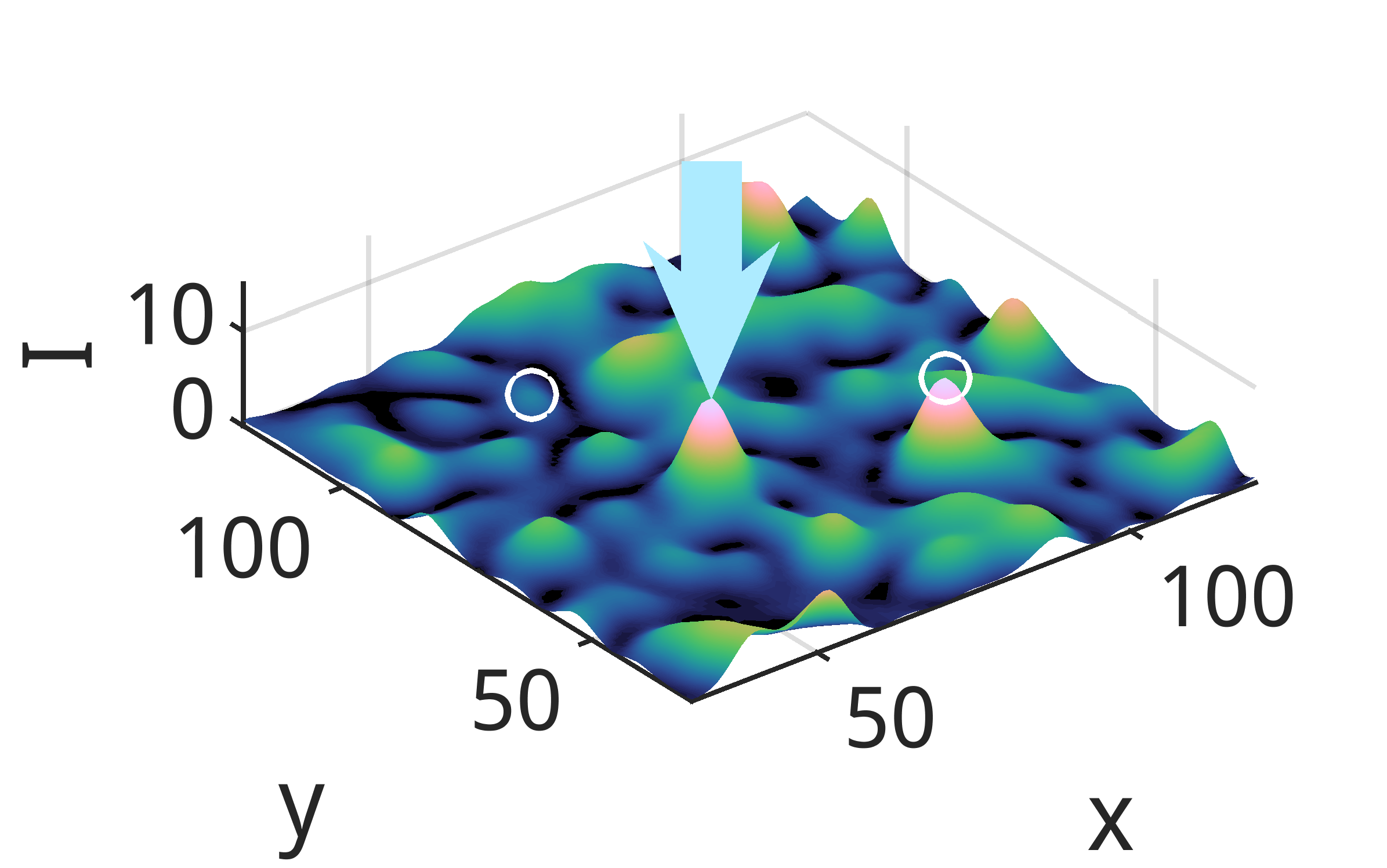}
  }
  
  \caption{Method for the individuation of the spatio-temporal maxima.
(a)-(c): three successive snapshots (separated by 9 ps) of a zoom on the field intensity transverse profile in the turbulent regime, for $\mu=5$ and $r=1$. The white circles indicate the spatio-temporal maxima detected with our method. Not all the spatial maxima have a circle in the image; each local spatial maximum (see for example the one indicated by the arrows) is followed during its time evolution: when its intensity is still growing (a) or is diminishing (c) in time (light blue arrows), it is not selected. The \textquotedblleft event" is counted only when the local spatial maximum reaches its maximum value in time (b, orange arrow).}
  \label{turbulent}
\end{figure}

The time averaged maximum intensity of the turbulent state is displayed in Fig. \ref{stationarystate} as a function of $\mu$, where for comparison we also show the intensity of the unstable homogeneous stationary solution, the maximum intensity of the stationary cavity solitons and the time averaged maximum intensity of chaotic solitons: the turbulent branch lies well above the other curves. Typical spatial profiles of the turbulent state are displayed in Figs. \ref{turbulent2}(a) and \ref{turbulent}.

For most of the simulations shown here (unless stated otherwise), we used the same set of parameters as in \cite{moving,VahedPRSA2014}, in particular we set $b=0.01$, $\alpha=2$, $\beta=1$, $\gamma=2$, $s=1$, $B=0.1$, and varied $r$ and $\mu$ as control parameters.

With respect to \cite{moving,VahedPRSA2014} we set here $\delta=0.01$ instead of zero, which amounts to having a complex coefficient in front of the Laplacian, that accounts for both diffraction and diffusion of the electric field. The additional diffusive term has been introduced phenomenologically, and, as stated above, it accounts for the finite linewidth of gain in absence of an equation for the material polarization.

Such a diffusive term is irrelevant as long as one deals with localized structures such as the stationary, oscillating or chaotic solitons of \cite{VahedPRSA2014} but it must be introduced in presence of an extended turbulent state because it acts as a filter for high spatial frequencies and prevents from the formation of filaments. Without that term the spatial structures contract rapidly and become very narrow intensity peaks with a flat Fourier spectrum, because energy is transported from the most unstable (low) wave vectors to the higher ones. Such narrow peaks cannot be sufficiently sampled over the numerical grid, and the occurrence of this self-collapse makes the simulations unreliable.

The stabilizing effect of the field diffusion term can be appreciated in the snapshot of the (transverse) spatial optical Fourier spectrum of the electric field, shown in figure \ref{turbulent2}(b). The spectrum is broad, showing the repartition of energy on many different spatial scales, but the size remains finite, and self--collapse is avoided.

\section{Extreme events} \label{extremeevents}
The method that we adopted to select the spatio-temporal maxima is illustrated in Fig. \ref{turbulent}. The local spatial maximum indicated by the light blue and orange arrows is not selected as long as it is growing in time and neither it is selected when its intensity is diminishing: it is selected only at the precise instant when it reaches its maximum in time (orange arrow). This procedure is applied to each local maximum of the spatial pattern throughout all the duration of the simulation: all the white circles in Fig. \ref{turbulent} indicate spatio-temporal maxima detected with this method.

The statistical analysis is done on all the spatio--temporal maxima recorded in this way during simulations lasting 25 ns, where we register one image of the transverse intensity distribution every ps, while the integration time-step is 100 fs. The spatial size of the integration window is $256 \times 256$ pixels, corresponding to about $256 \times 256$ $\mu$m (the spatial step used being 0.25).

We used three different definitions for the threshold that determines whether an event may be regarded as extreme.

Threshold 1: the mean intensity, averaged on every point of the transverse plane and every instant in time, plus 8 times the standard deviation.
This is the definition most commonly used for studying optical rogue waves in spatially extended systems \cite{arecchi,oppo,liu,oppoPRL2016}.

Threshold 2: two times the significant wave height $H_s$, defined as the average of the highest third of the spatio-temporal maxima values. This is the typical hydrodynamic definition, and permits to get rid of a possible global increase of the average value, that would not correspond to a freak wave. Note that due to the large number of very low--intensity peaks, which would make the treatment and data analysis uselessly time--consuming, we computed the significant wave height $H_s$ excluding events whose height is smaller than a given threshold, which is about $0.5$. This cut makes more stringent the criterion for the definition of extreme events. The typical number of remaining \textquotedblleft events" detected during a simulation is around $6\times 10^{5}$.

Threshold 3: average of spatio-temporal maxima values plus 8 times the standard deviation. This is a new definition, proper to our method and it is, by far, the most stringent one. We decided to introduce this third threshold because it is the equivalent of threshold 1, most commonly used, but it is most appropriate for our numerical data, representing the spatio--temporal maxima obtained with our method.
\begin{figure}[t!]
  \centerline{\subfigure[ ]{\includegraphics[width=0.5\columnwidth]{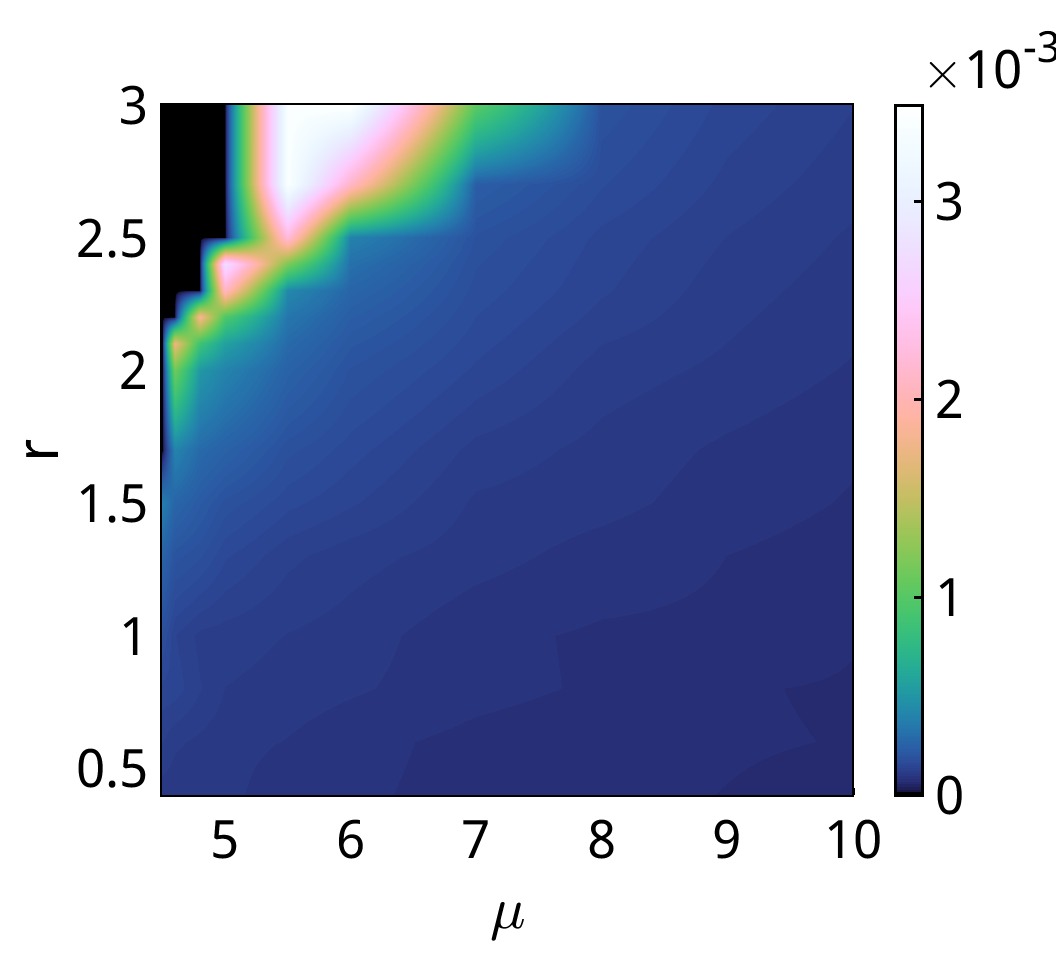}
    }
      \hfil
    \subfigure[ ]{\includegraphics[width=0.5\columnwidth]{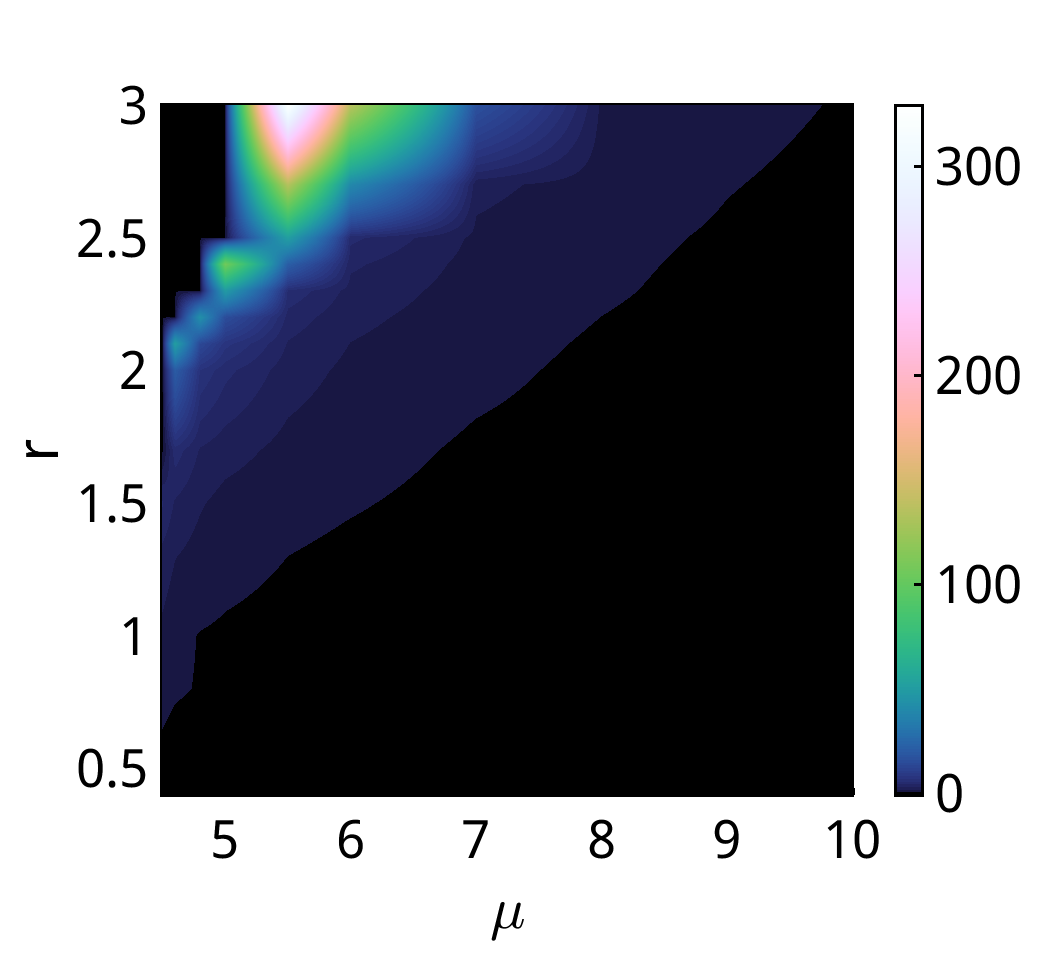}
  }}
  \caption{Density plots showing: (a) fraction of rogue events using threshold 1 and (b) excess kurtosis of the total intensity PDF calculated with respect to that of the negative exponential ($K_\mathrm{exp}=9$), as a function of the parameters $r$ and $\mu$.}
  \label{density_plots_int}
\end{figure}
\begin{figure}[t!]
  \centerline{\subfigure[ ]{\includegraphics[width=0.5\columnwidth]{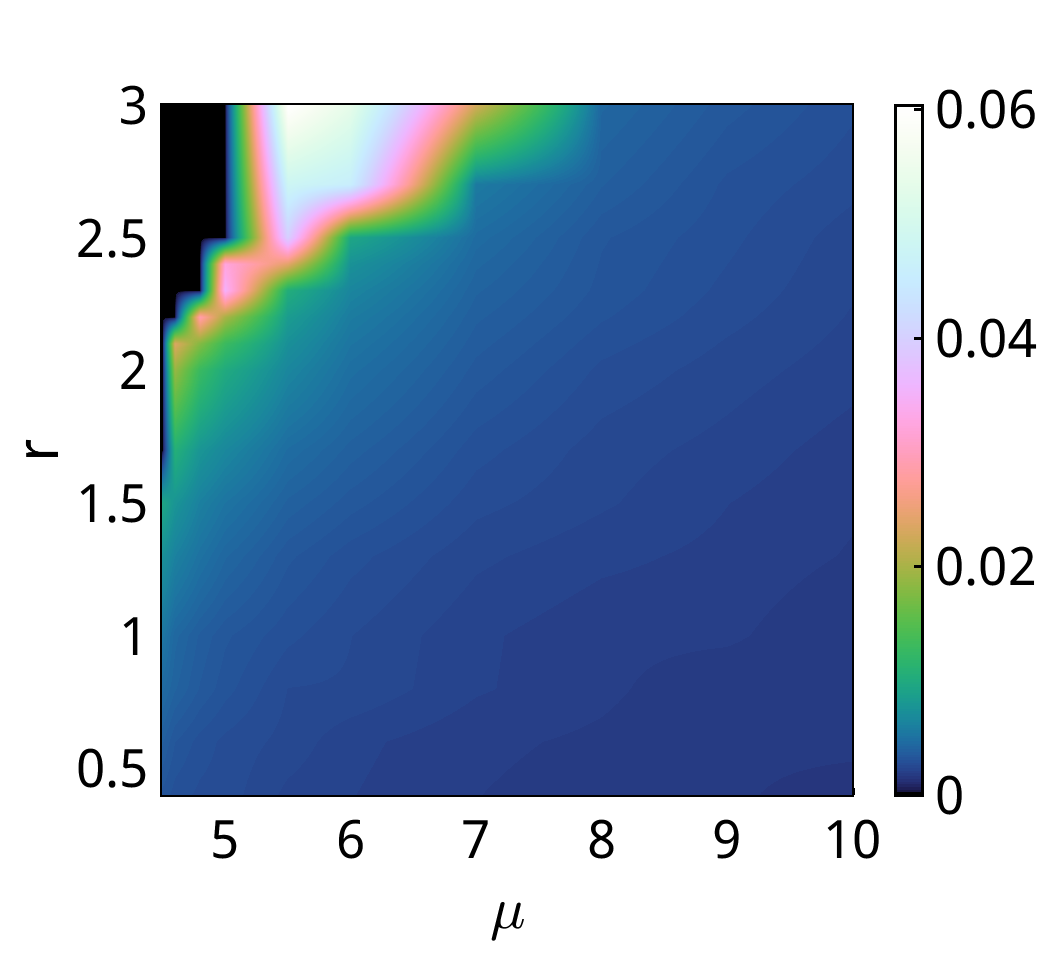}
    }
      \hfil
    \subfigure[ ]{\includegraphics[width=0.5\columnwidth]{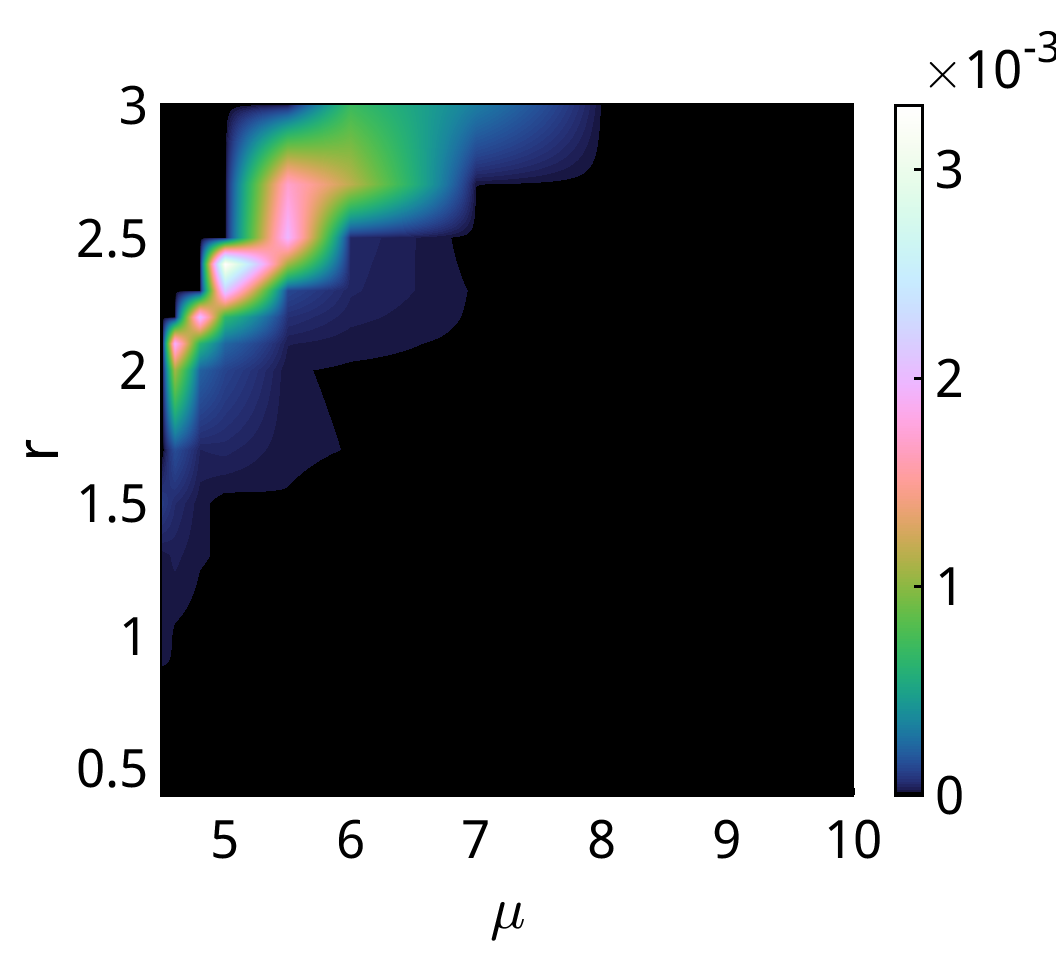}
  } }
    \subfigure[ ]{\includegraphics[width=0.5\columnwidth]{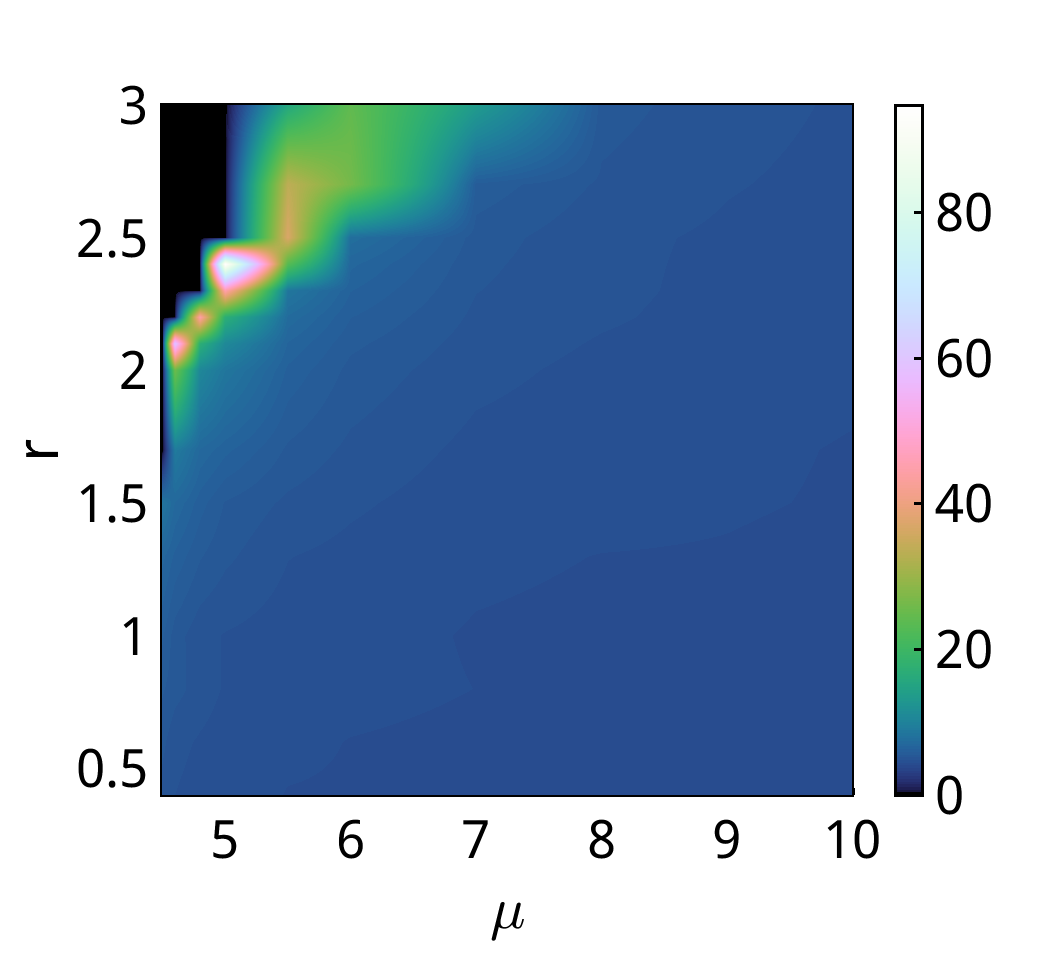}
  }
  \caption{Density plots showing: (a) fraction of rogue events using threshold 2 and (b) threshold 3, and (c) kurtosis of the PDF of the spatio--temporal maxima as a function of parameters $r$ and $\mu$ (as a reference, for a Gaussian distribution $K_\mathrm{Gauss}=3$).}
    \label{density_plots_max}
\end{figure}

\section{Dependence on laser parameters} \label{parameters}
We performed numerical simulations for different values of the control parameters $\mu$ (pump parameter) and $r$ (ratio of carrier lifetimes) to determine under which conditions rogue waves are more likely to be observed in an experiment. As indicators for the rogue nature of the data we used: i) the ratio of the number of extreme events (identified according to the three thresholds defined above) to all the spatio-temporal maxima, and ii) the kurtosis of the data distribution, which is the ratio of the fourth moment about the mean to the square of the variance.

We display the results of the simulations \textit{via} colorscale density plots of the two indicators.
Fig. \ref{density_plots_int} refers to the statistics made on \textit{all the intensity values} and shows the fraction of rogue waves according to threshold 1 (a) and the kurtosis of the PDF (b) with respect to that of a negative exponential
\begin{equation}
\frac{1}{\langle I\rangle}\exp\left(-\frac{I}{\langle I\rangle}\right)\,,
\end{equation}
which is the PDF corresponding to a Gaussian statistics on the field amplitude.

Fig. \ref{density_plots_max} refers to the statistics made only on the \textit{spatio-temporal maxima} and shows the fraction of rogue waves according to thresholds 2 (a) and 3 (b) and the excess kurtosis with respect to that of a Gaussian distribution for $I_\mathrm{max}$ (c).

All these data plots are visually similar, showing a typical structure for the maximum extreme events probability placed approximately at the left boundary of the turbulent branch, but with some small differences that are peculiar to the different quantity in study.

\begin{figure}
\includegraphics[width=0.8\linewidth]{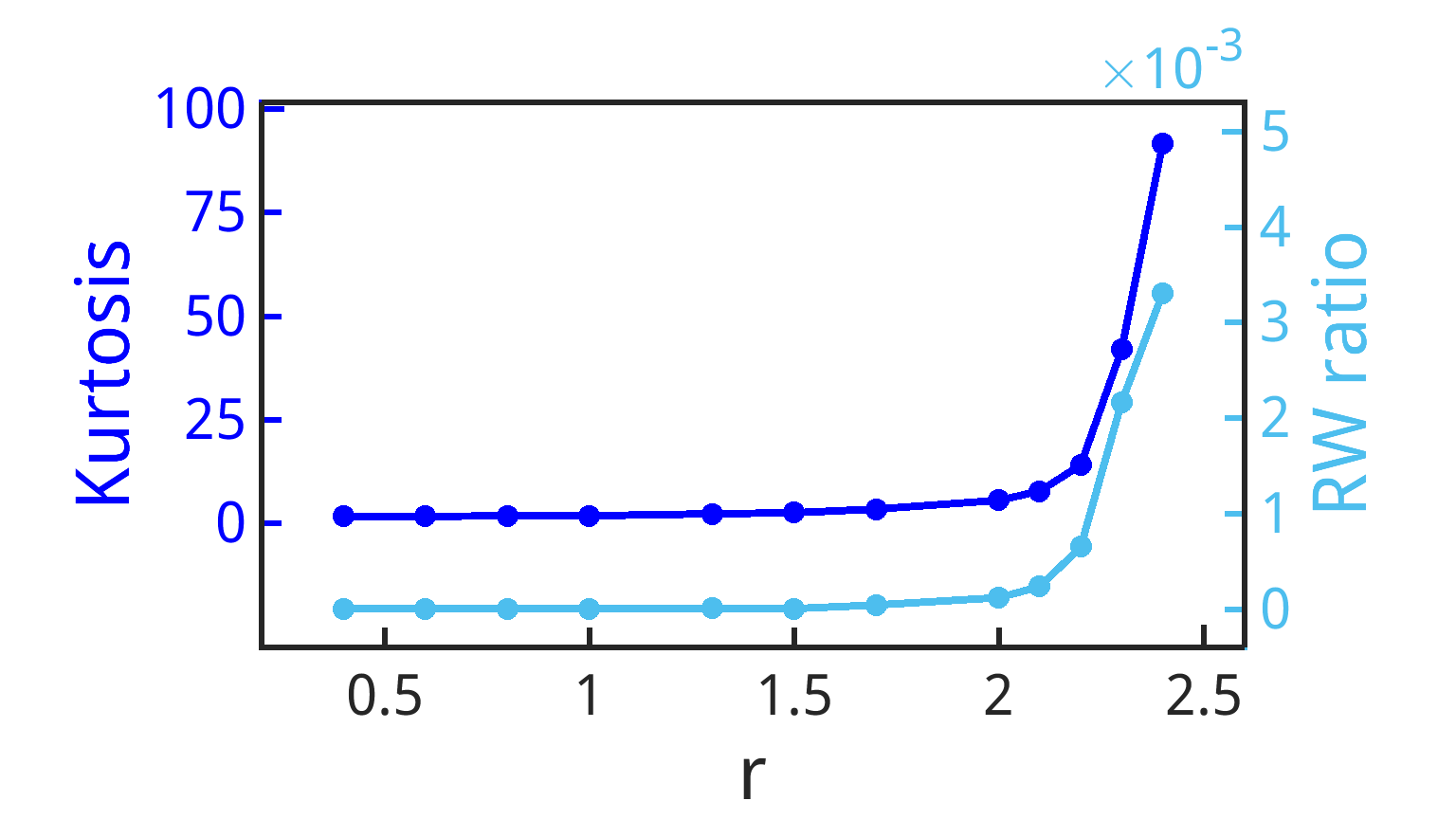}
\includegraphics[width=0.8\linewidth]{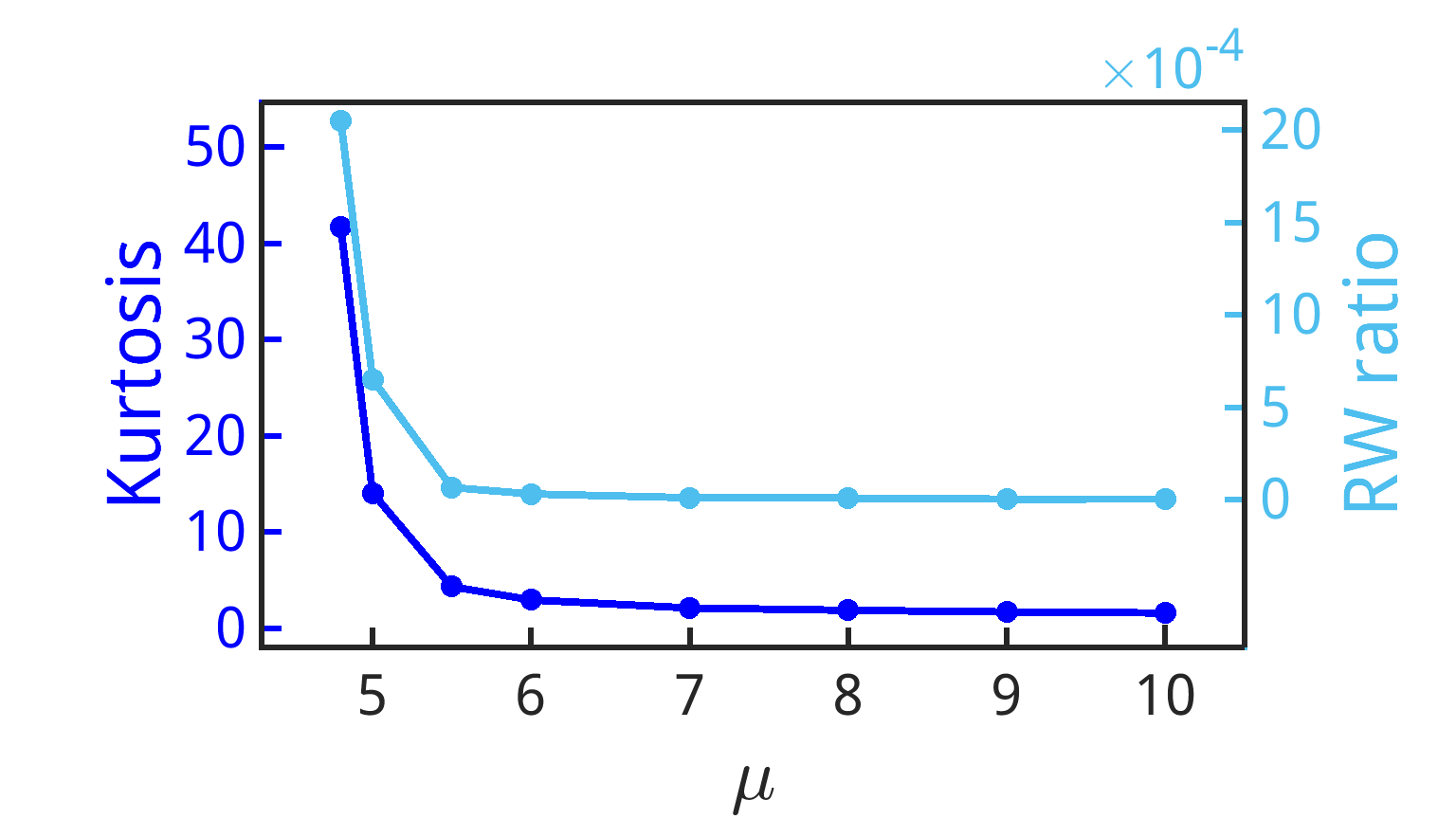}
\caption{Fraction of rogue waves according to threshold definition 3 (light blue, right vertical axis) and kurtosis of the PDF of the spatio--temporal maxima (blue, left vertical axis), as a function of $r$ for $\mu=5$ (a) and as  a function of $\mu$ for $r=2.2$ (b).}
\label{kurtrogue}
\end{figure}

Figs. \ref{kurtrogue}(a,b) show two sections of Figs. \ref{density_plots_max}(b,c) for fixed $\mu=5$ (below the laser threshold) and variable $r$ (a) and for fixed $r=2.2$ and variable $\mu$ (b). Both indicators show a rapid increase in correspondence with the maxima shown in the density plots.

\begin{figure}[htb!]
  \centering{\subfigure[ ]{\includegraphics[width=0.9\columnwidth]{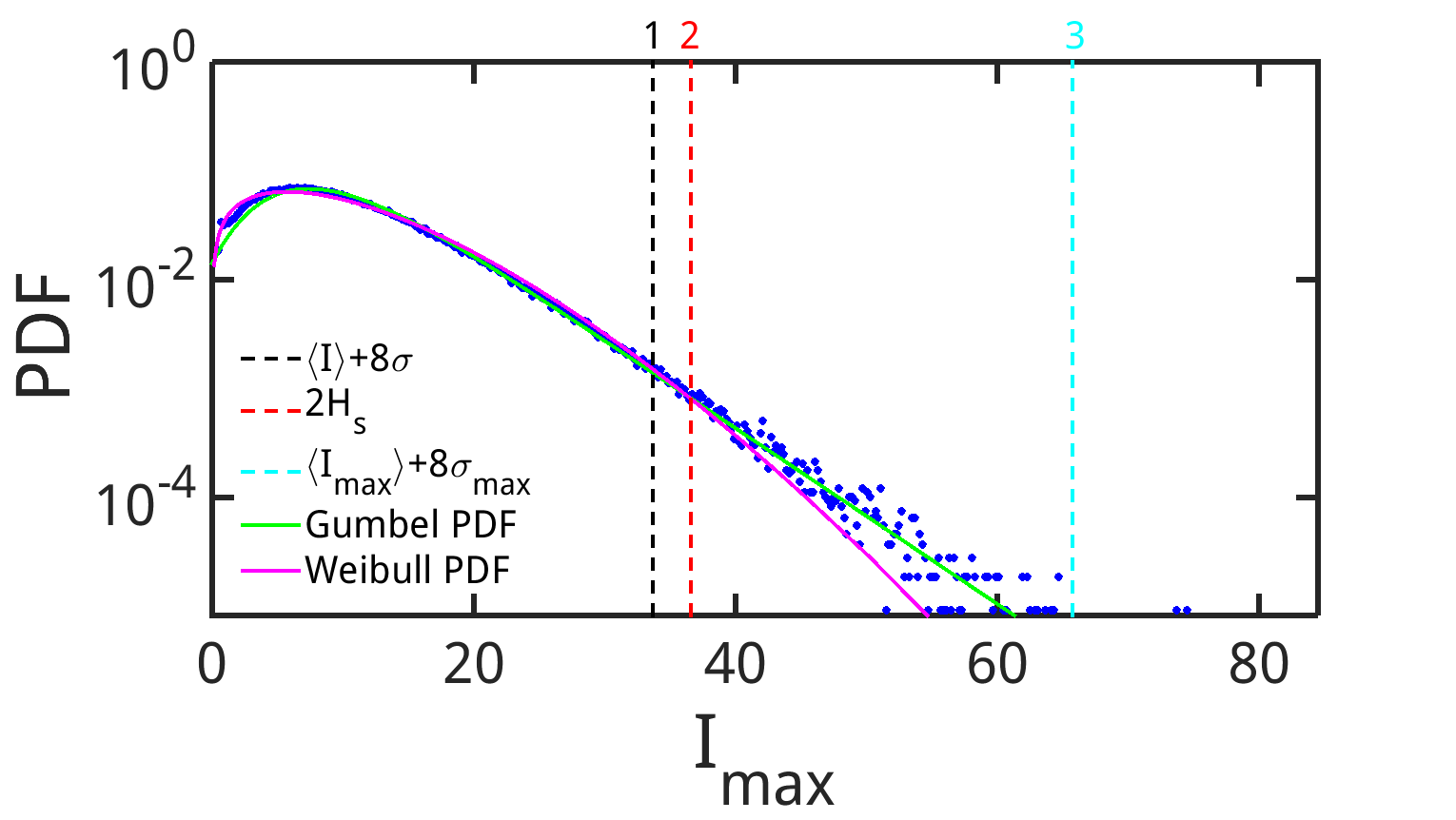}
    }
      \hfil
    \subfigure[ ]{\includegraphics[width=0.9\columnwidth]{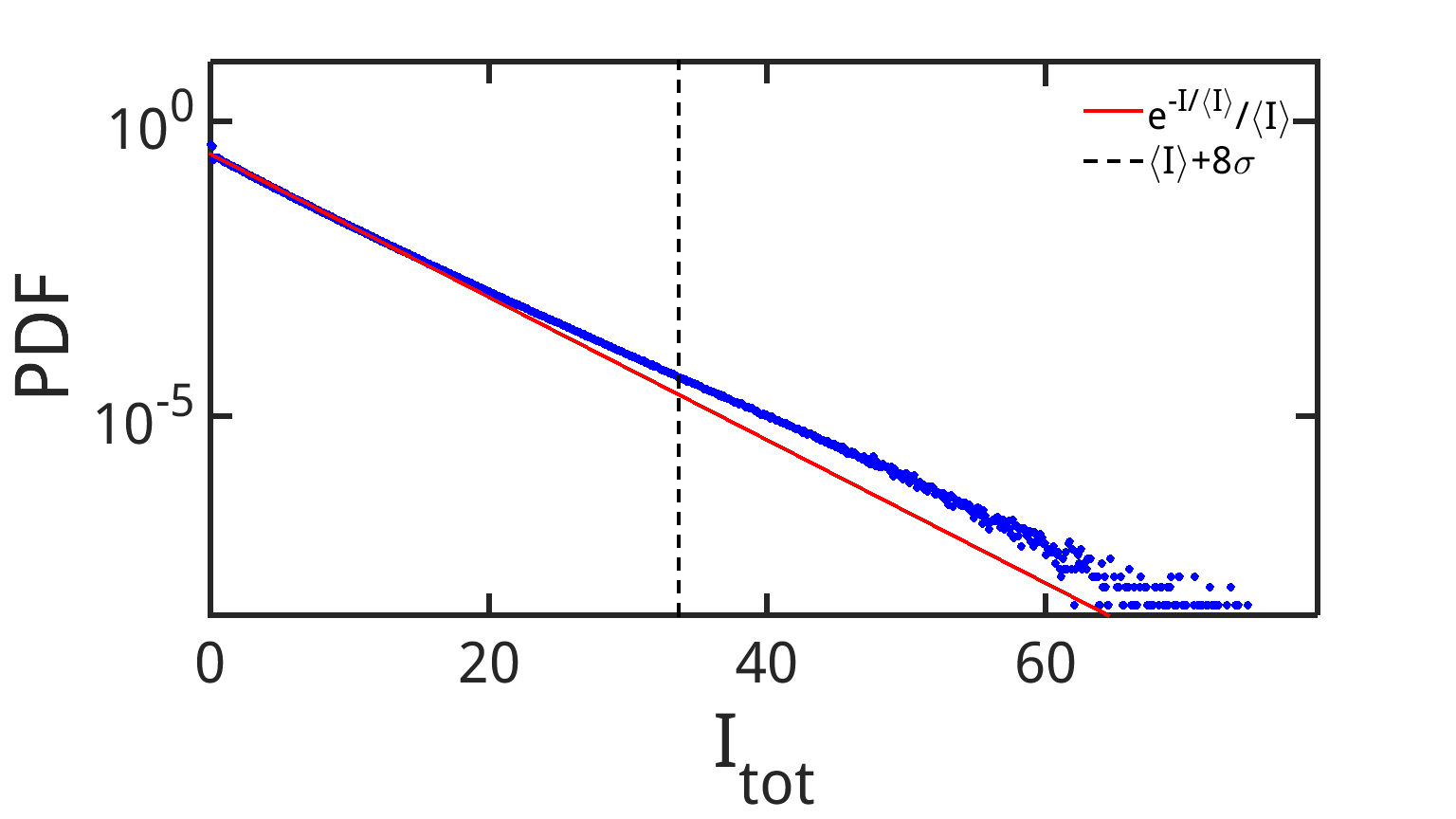}
  }
  }
  \caption{(a) Probability density function (PDF) of all the spatio-temporal maxima detected with our method during a numerical simulation lasting 25 ns, for $\mu=7$ and $r=2.5$. The green and magenta lines are respectively the Gumbel and Weibull distributions computed from the mean and standard deviation of the data. The three vertical dashed lines indicate three different definitions of rogue wave thresholds, defined in the text as thresholds 1,2 and 3 (see the legend, and the text).
(b) Probability density function (PDF) of all the values explored by the intensity during the entire simulation in each point of the transverse plane. Black dashed vertical line: threshold for rogue waves (same as threshold 1 in (a)).}
  \label{stat1}
\end{figure}
\begin{figure}[t!]
  \centering{\subfigure[ ]{\includegraphics[width=0.9\columnwidth]{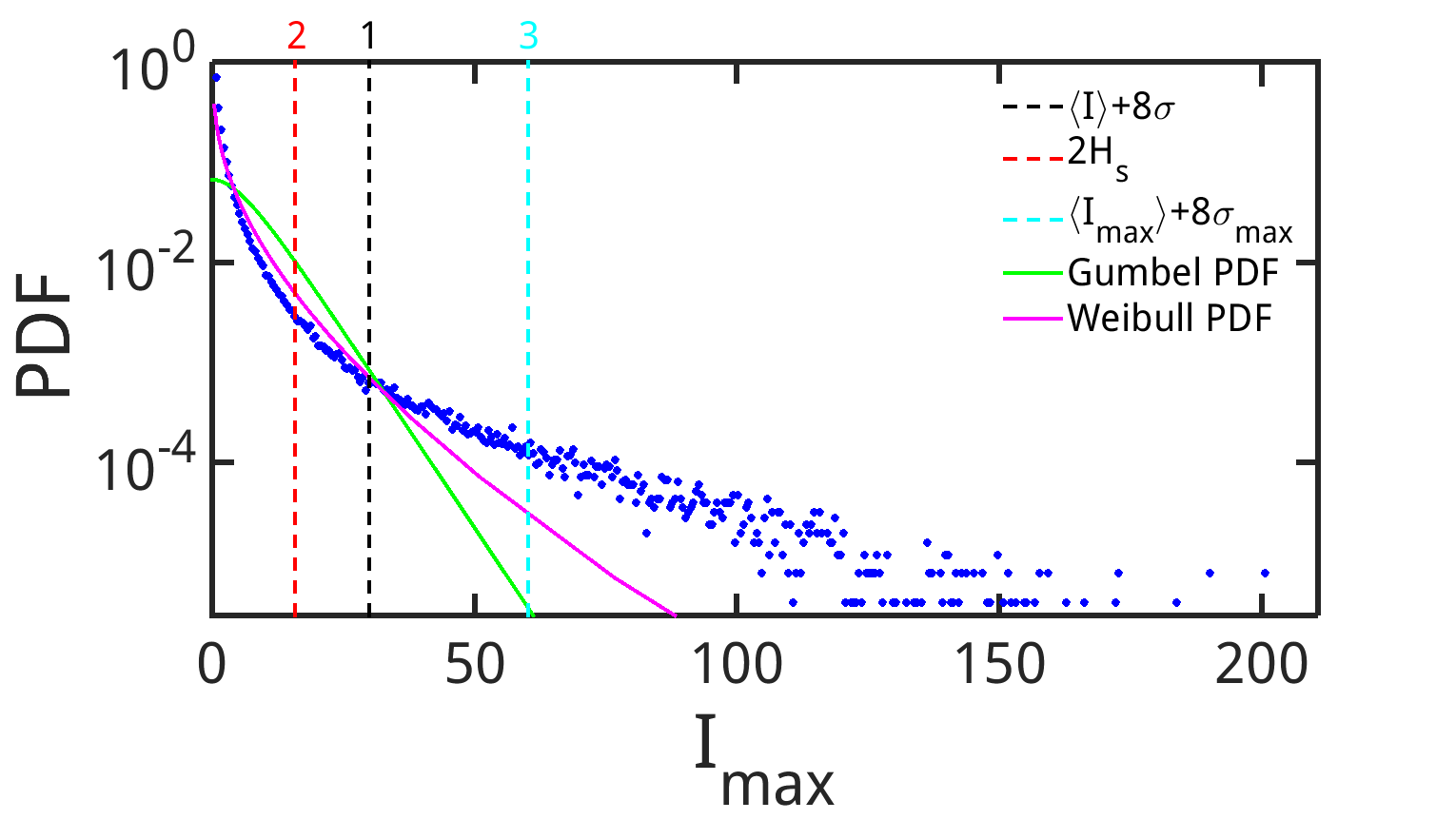}
    }
      \hfil
    \subfigure[ ]{\includegraphics[width=0.9\columnwidth]{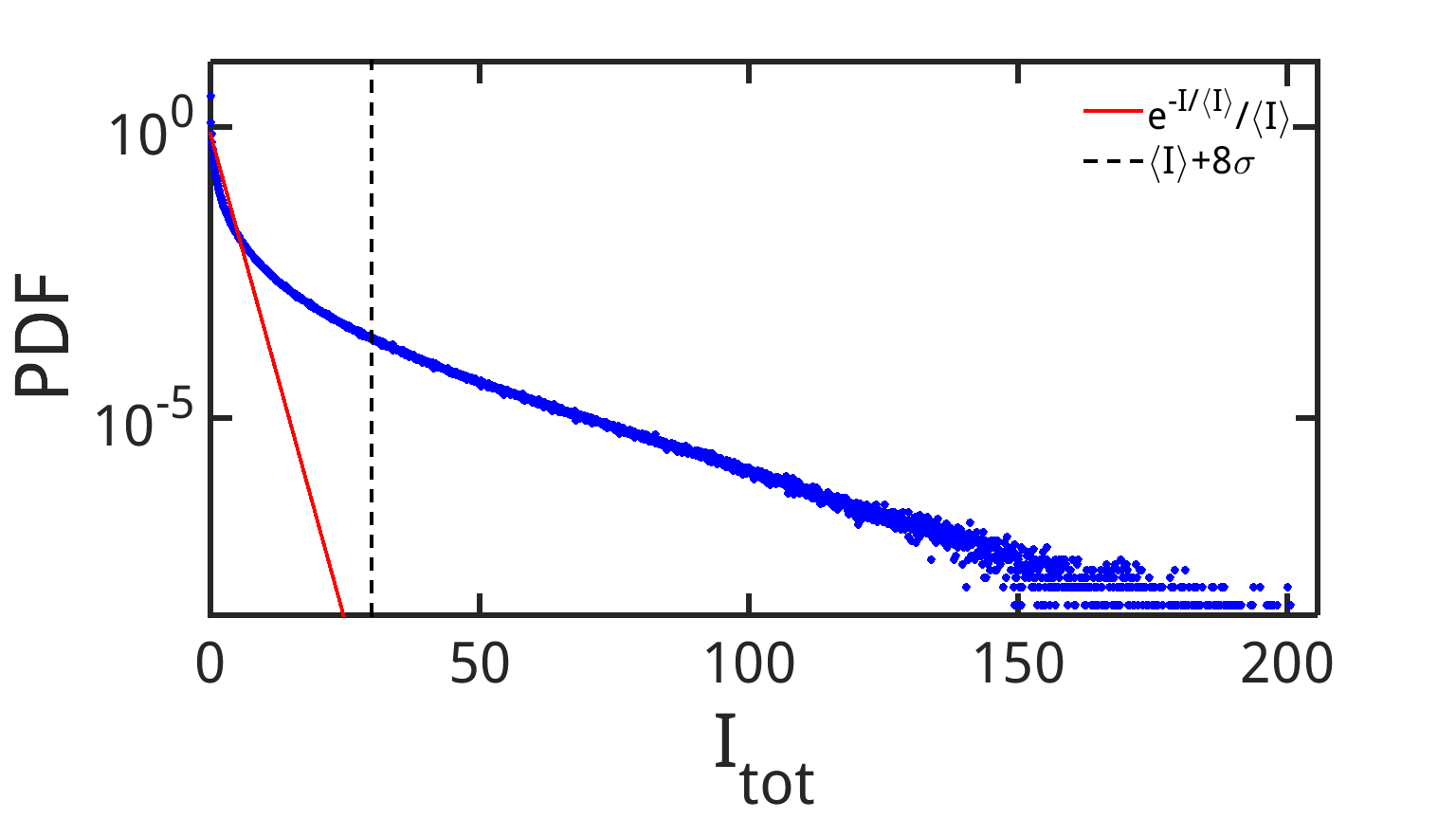}
  }
  }
  \caption{(a) and (b): same plots as in figure \ref{stat1}, but for $\mu=5$ and $r=2.4$. The presence of very heavy tails is clearly visible, and RW exist according to all the threshold definitions.}
  \label{stat2}
\end{figure}

We can therefore conclude that rogue waves are most probable for low pump $\mu$, below the laser threshold, where the turbulent state coexists with the non-lasing solution and the localized structures, and for high values of $r$, corresponding to a fast absorber which favors a Q-switching-like behaviour \cite{VahedPRSA2014}.

Fig. \ref{stat1}(a) shows the probability density function (PDF) of the spatio-temporal maxima for $\mu=7$ and $r=2.5$, i.e. for a set of parameters
for which the simulations show that the probability of extreme events is small.

The three vertical dashed lines indicate the three threshold defined above. While according to threshold 1 and 2 there is a considerable number of extreme events, only very few events lie beyond threshold 3. The data follows well the behavior predicted by the generalized extreme value (GEV) theory.
In particular, the distributions that better describe the behaviour of our data are provided by the Gumbel distribution
\begin{equation}
\frac{1}{\beta}\exp{[-(z+\exp{(-z)})]}\,,\quad z=\frac{I_{max}-\langle I_{max}\rangle}{\beta}+\gamma\,,
\end{equation}
where $\beta$ is a fitting parameter and $\gamma$ is Euler's constant, and the  Weibull distribution
\begin{equation}
\frac{k}{\lambda}\left(\frac{I_{max}}{\lambda}\right)^{k-1}\exp{\left[-\left(\frac{I_{max}}{\lambda}\right)^k\right]}\,,
\end{equation}
where $k$ and $\lambda$ are fitting parameters. The Weibull distribution, however, predicts a slightly more rapidly decaying tail.

Fig. \ref{stat1}(b) shows instead the PDF of all the intensity values which displays a small but clear deviation from the negative exponential.

Fig. \ref{stat2} is the same as Fig. \ref{stat1} but for a most favorable case ($\mu=5$ and $r=2.4$) for RW existence. Here the Gumbel and Weibull distribution do not approximate well the PDF of the spatiotemporal maxima anymore, and a large number of extreme events exists according to all three thresholds. Also, Fig. \ref{stat2}(b) shows a more pronounced deviation from the negative exponential than Fig. \ref{stat1}(b).

\begin{figure}
\includegraphics[width=1\linewidth]{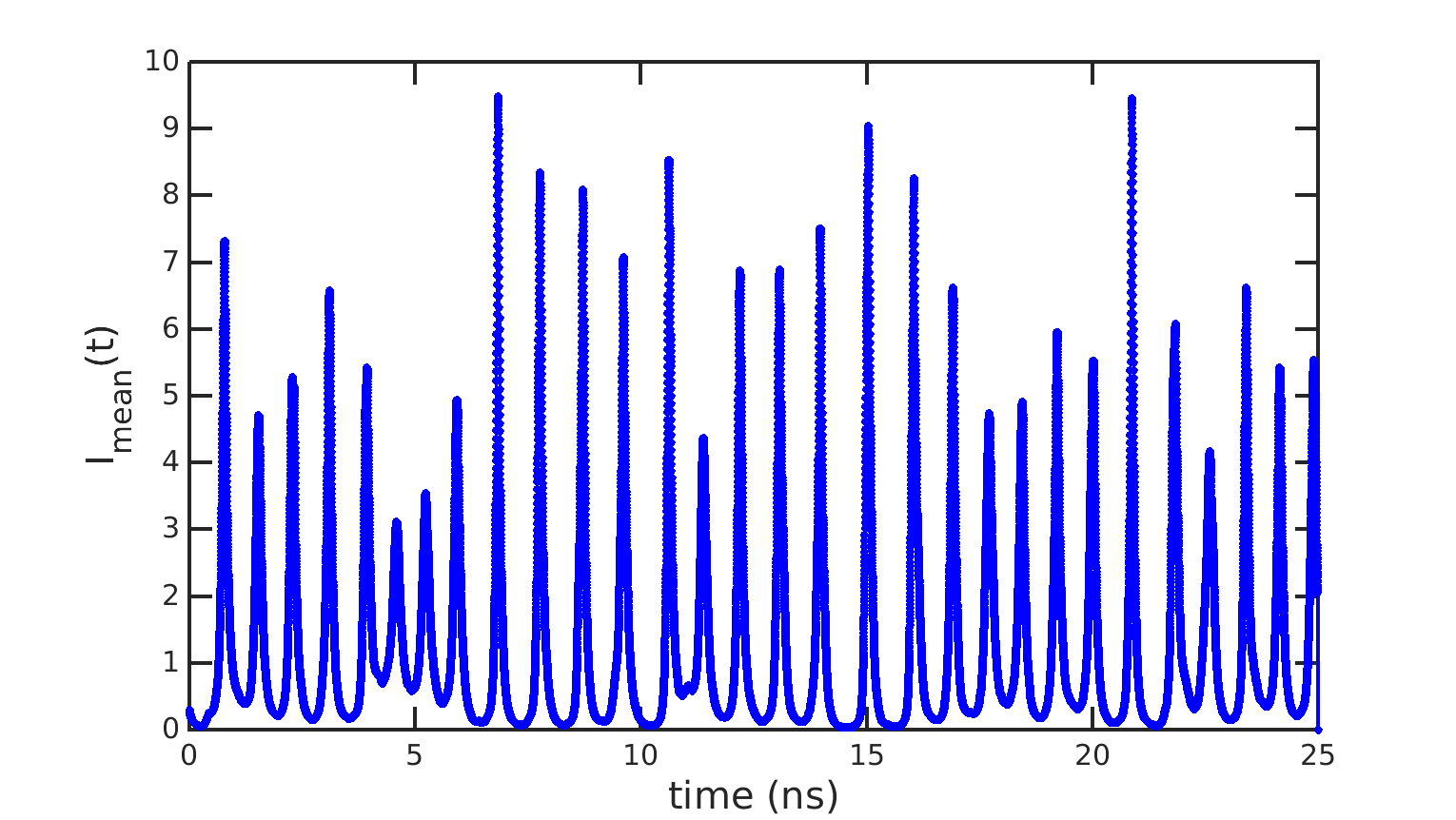}
\caption{Temporal evolution of the spatially averaged intensity for the same simulation as in Fig. \ref{stat2}.}
\label{stat3}
\end{figure}

When looking at the temporal evolution of the spatial intensity profile $I(x,y)$, one can observe that in this case the turbulent state shows a global superimposed rather regular oscillation between almost null intensity and a maximum intensity, as clearly shown in Fig. \ref{stat3}, where we plotted the temporal evolution of the spatial averaged field intensity. This is probably a residual effect of the Q--switching instability that affects the system for high values of $r$, both in the plane--wave case, where the dependence on $(x,y)$ of the electric field is neglected, and in presence of cavity solitons.
\begin{figure}[t!]
  \centering{\subfigure[ ]{\includegraphics[width=0.9\columnwidth]{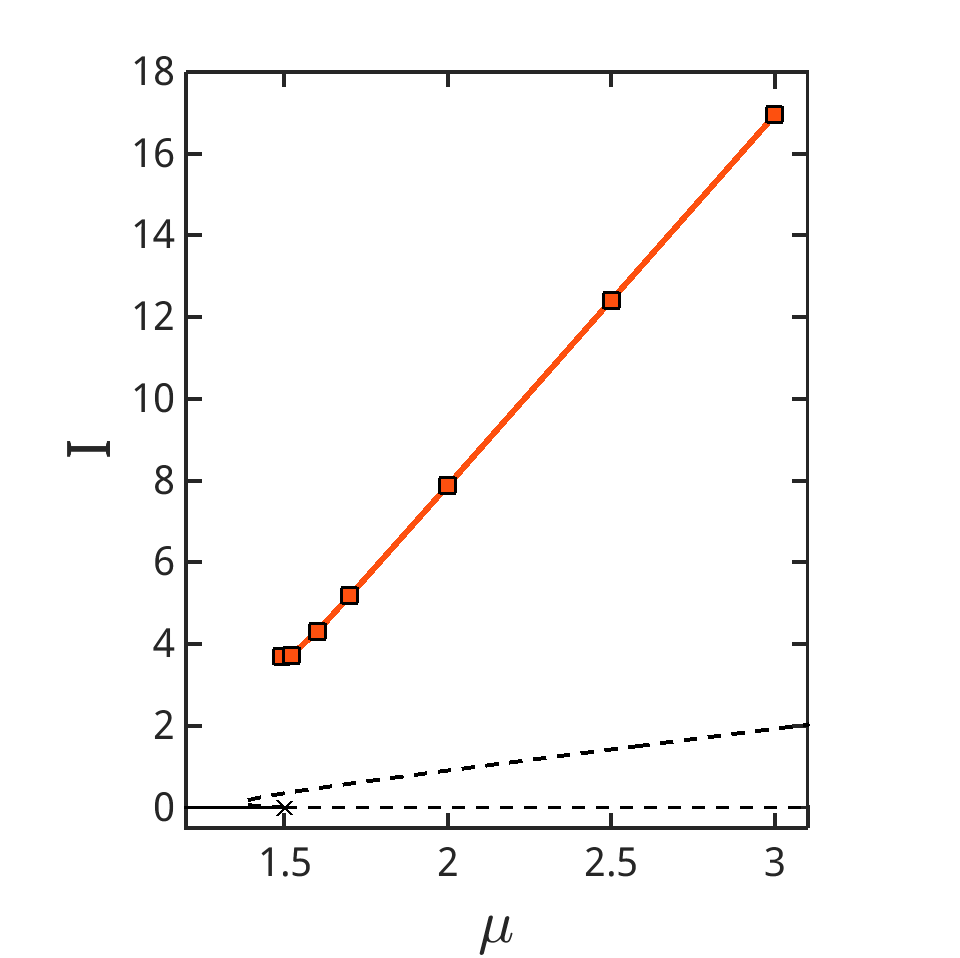}
    }
      \hfil
    \subfigure[ ]{\includegraphics[width=0.9\columnwidth]{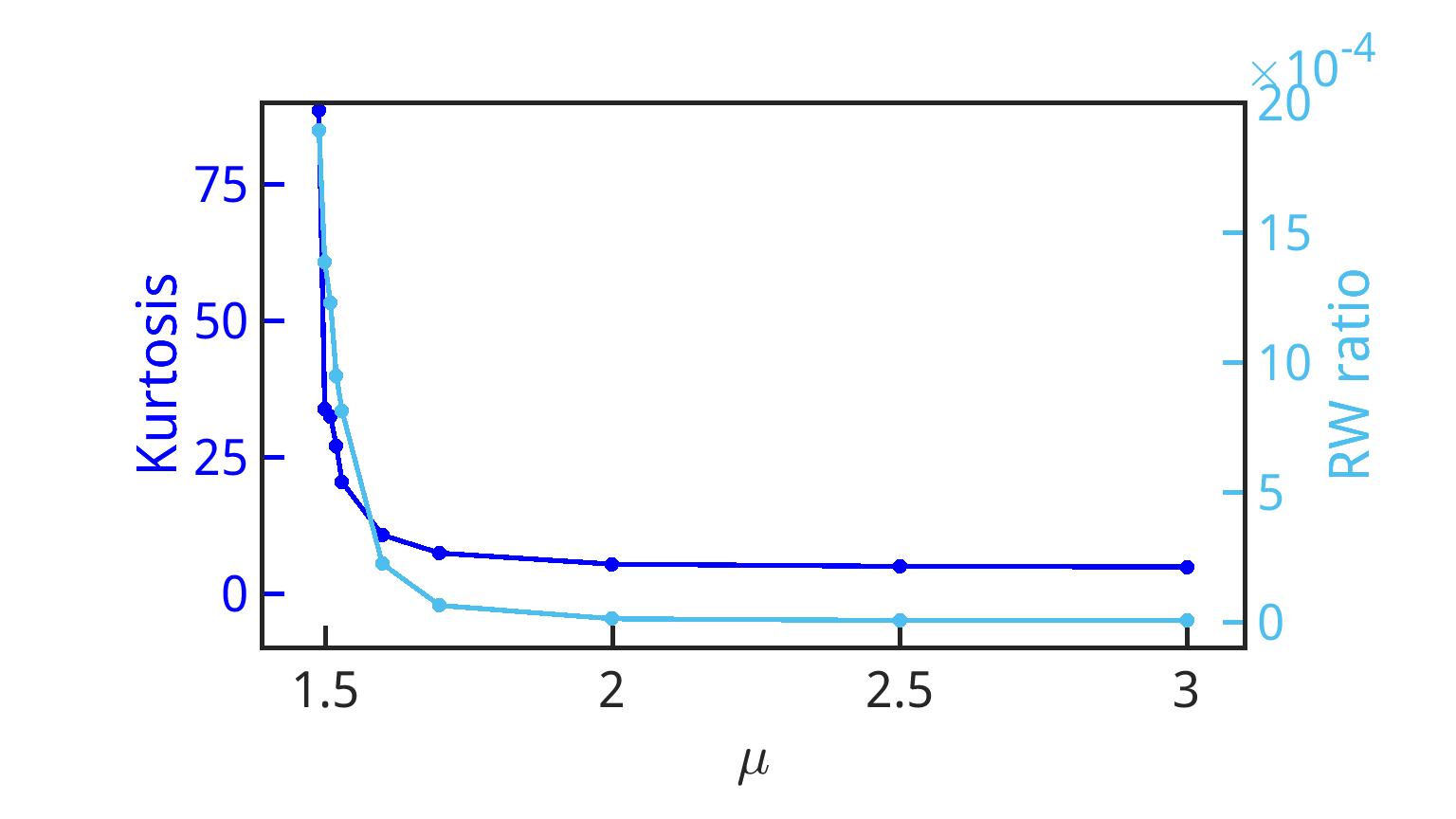}
  }
  }
  \caption{(a) Unstable stationary homogeneous solution (black dashed line) and turbulent branch (red line with symbols) for the parameters $r=1$, $b=0.005$, $\alpha=2$, $\beta=0$, $\gamma=0.5$, $s=10$, $B=0.$, $\delta=0.01$. The laser threshold is at $\mu_{th} = 1.5$ (b) Kurtosis of the PDF of the spatio--temporal maxima (blue, left axis) and fraction of rogue events using threshold definition 3 (light blue, right axis), as a function of parameter $\mu$, for simulations lasting $25$ ns. }
  \label{barbay}
\end{figure}

We can compare our results with those of \cite{selmi}. In both cases rogue waves appear to be related to spatio-temporal complexity, but in \cite{selmi} the proportion of RW and the excess kurtosis of the data distribution seem to \textit{increase} (at least for a large set of the pump values above threshold) when the pump $\mu$ is increased, in contrast with our results, where both indicators \textit{decrease}.

It is important to remark that in \cite{selmi} the date analysed are those of the \textit{mean} intensity, averaged over the spatial integration window, which can present a completely different behaviour with respect to the local intensity. For instance in a situation such that of Fig. \ref{stat2}, where our RW indicators calculated on the spatio-temporal maxima are largest, the mean intensity is conversely very well-behaving, as shown in our Fig. \ref{stat3}.
A statistic analysis made on the spatially averaged intensity would probably show no trace of extreme events in this case.

To substantiate this interpretation we performed a new set of numerical simulations using the same parameters as in Ref. \cite{selmi}, that is $r=1$, $b=0.005$, $\alpha=2$, $\beta=0$, $\gamma=0.5$, $s=10$, $B=0.$, $\delta=0.01$, and performed the statistical analysis of the spatio--temporal maxima obtained with our method. 
In Fig. \ref{barbay} (a) we show the stationary homogeneous solution and the turbulent branch, while in (b) the Kurtosis of the PDF and the RW fraction are shown as a function of the pump parameter $\mu$ (same as in Fig. \ref{kurtrogue}). The same behavior as for all the other examples shown in this paper is present here: the maximum probability of extreme events occurs for low pump values, that is, at the left boundary of the turbulent branch, in contrast with the results of Ref. \cite{selmi}.

\section{Spatial and temporal profiles} \label{profiles}

Once observed the presence of rogue waves in the system under analysis, we studied the spatial and temporal profiles of the rogue waves detected during the numerical simulations.

\begin{figure}[t!]
  \centering{\subfigure[ ]{\includegraphics[width=0.85\columnwidth]{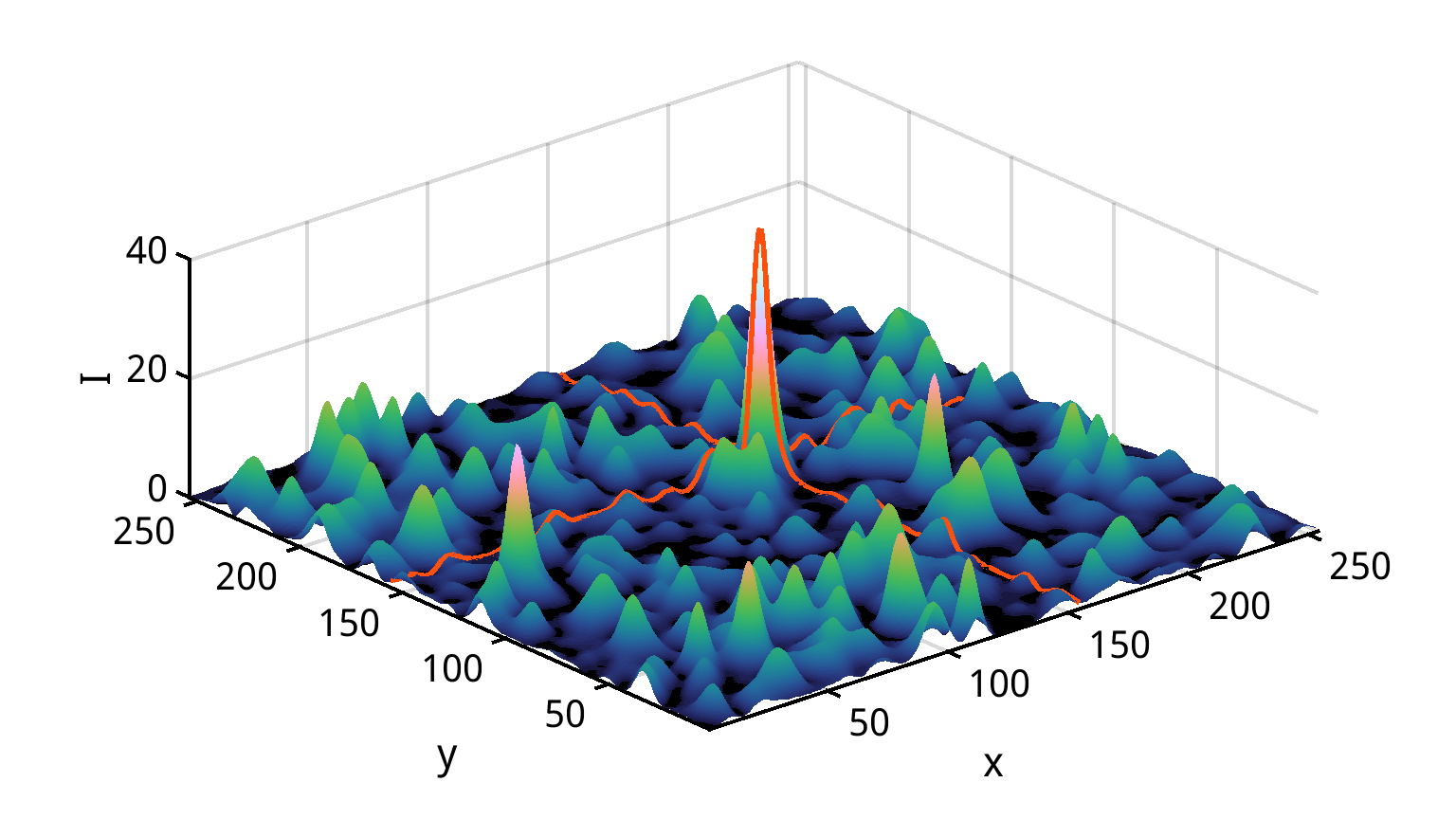}
    }
      \hfil
    \subfigure[ ]{\includegraphics[width=0.85\columnwidth]{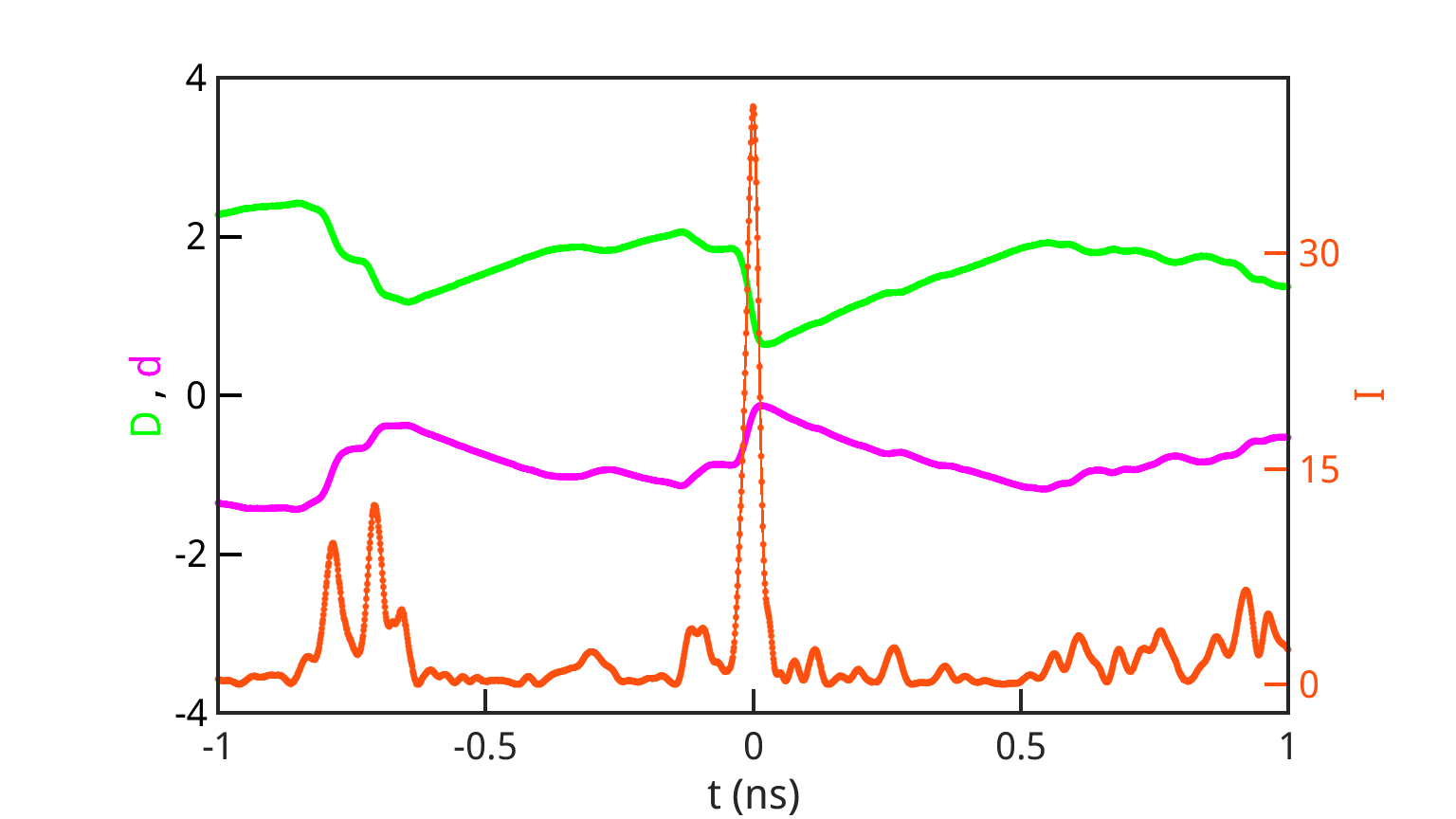}
  }
      \hfil
    \subfigure[ ]{\includegraphics[width=0.85\columnwidth]{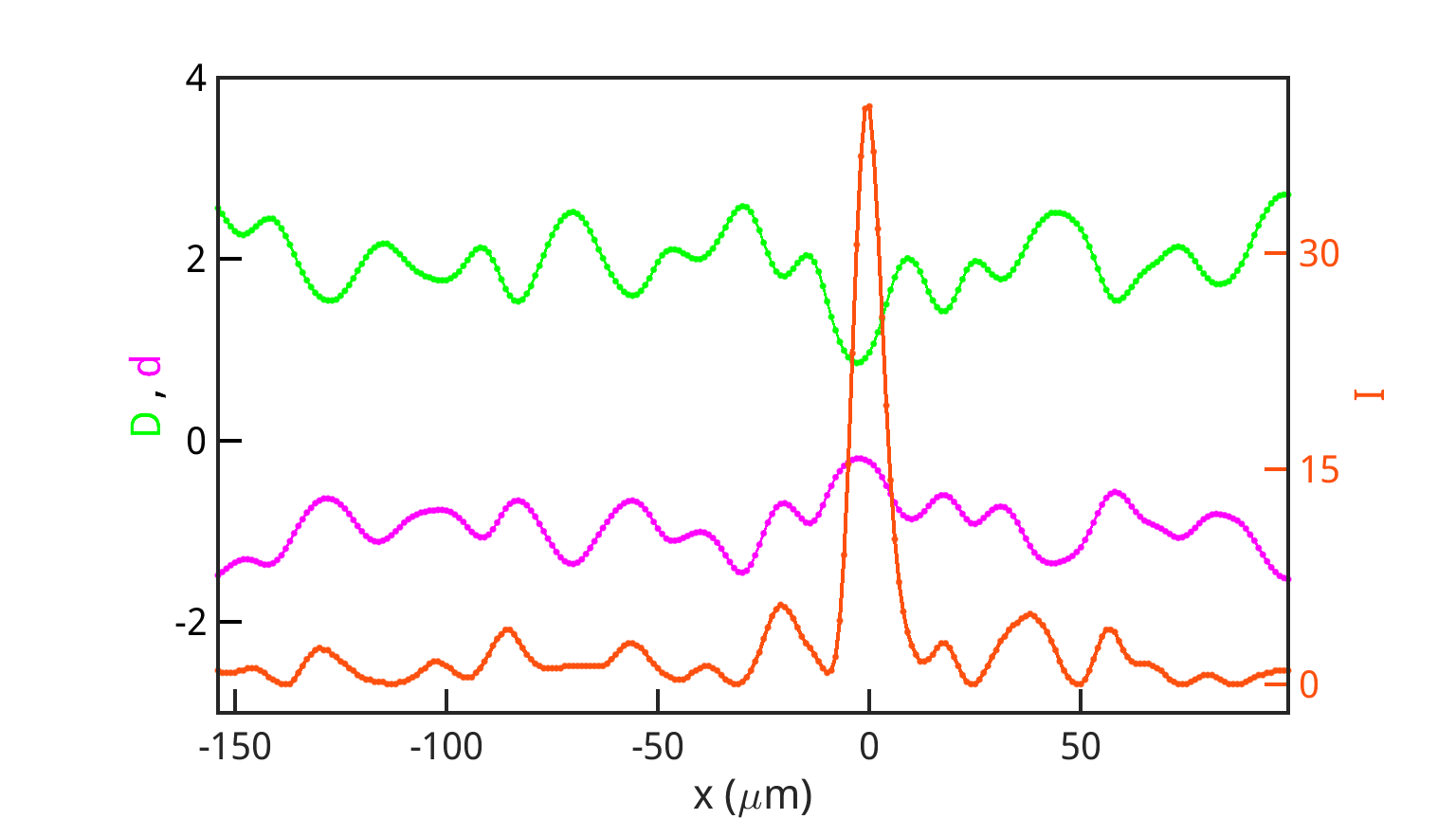}
  }
  }
  \caption{Example of a rogue wave in the transverse plane (a) and its temporal (b) and spatial (c) profiles, shown for the variables $I$, $D$ and $d$. Parameters are $\mu = 4.8, r = 2.2$.}
  \label{profile}
\end{figure}

In Fig. \ref{profile} we show an example of RW and its spatial and temporal profiles for the field intensity, and the carrier populations in the active and passive media.
For the sake of simplicity, we limited the spatial analysis to the $x$ and $y$-axis. The temporal profile is given by the values registered throughout the simulation in the spatial point where the rogue wave is detected. From the spatial and temporal profiles it is possible to get the minimal FWHM detected during each simulation.
The typical FWHM in time is 16 ps and the typical FWHM in space is 6 $\mu$m: as we noticed, these values remain almost constant throughout all the different simulations (performed with different values of $\mu$ and $r$), suggesting that there is a typical spatial and temporal size for this kind of phenomena.

\begin{figure}[t!]
  \centering{\subfigure[ ]{\includegraphics[width=0.85\columnwidth]{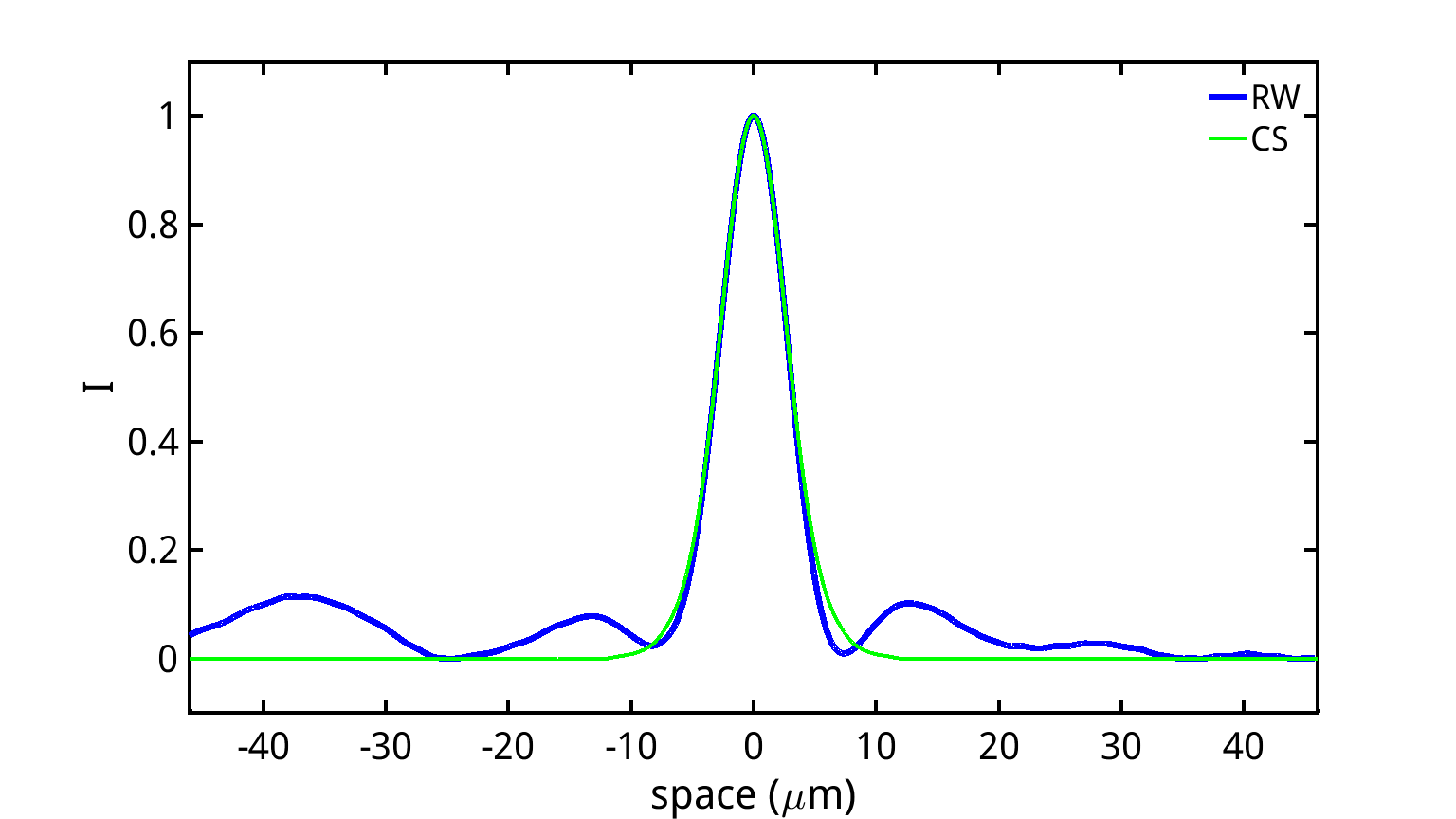}
    }
      \hfil
    \subfigure[ ]{\includegraphics[width=0.85\columnwidth]{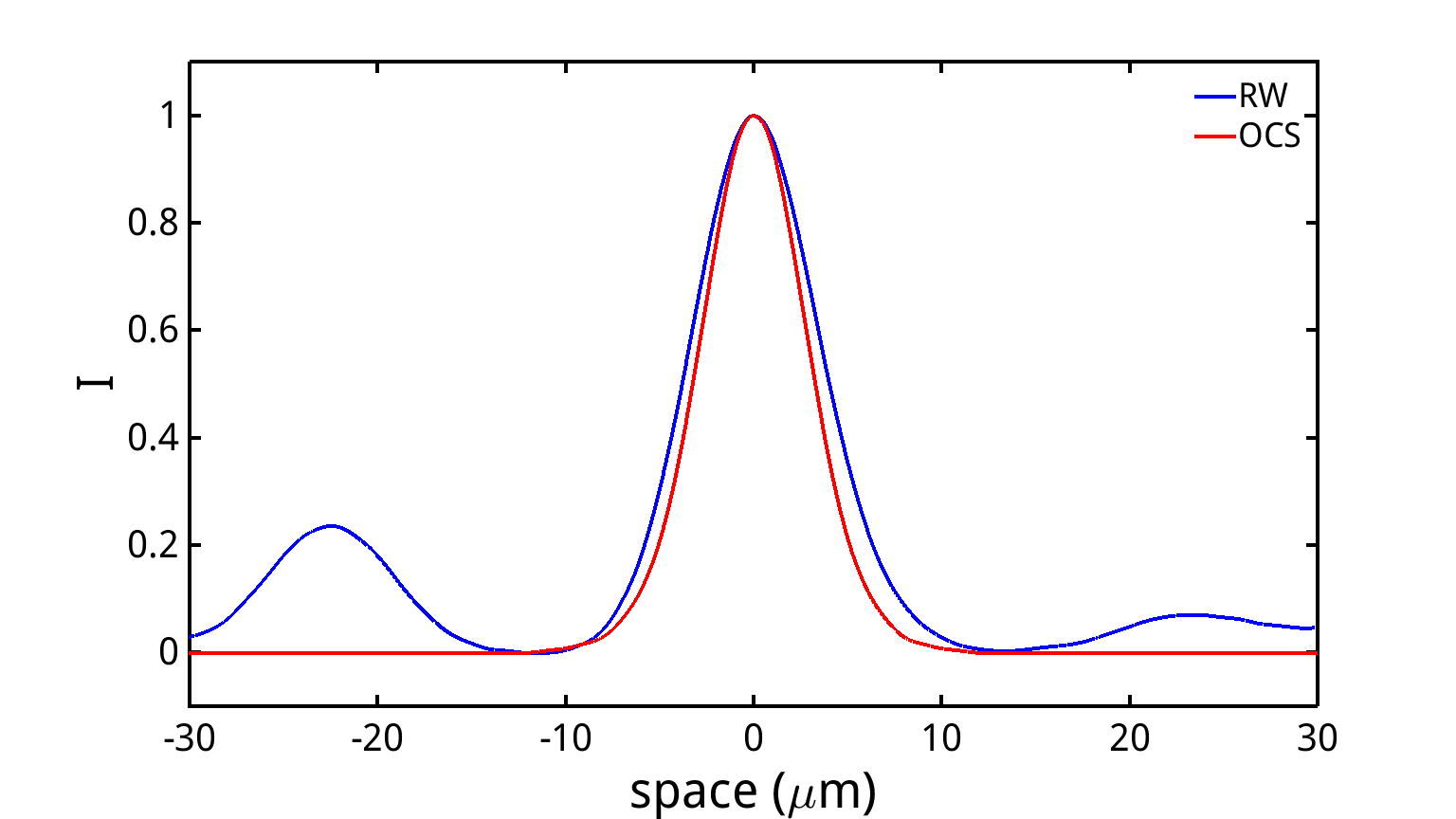}
  }
      \hfil
    \subfigure[ ]{\includegraphics[width=0.85\columnwidth]{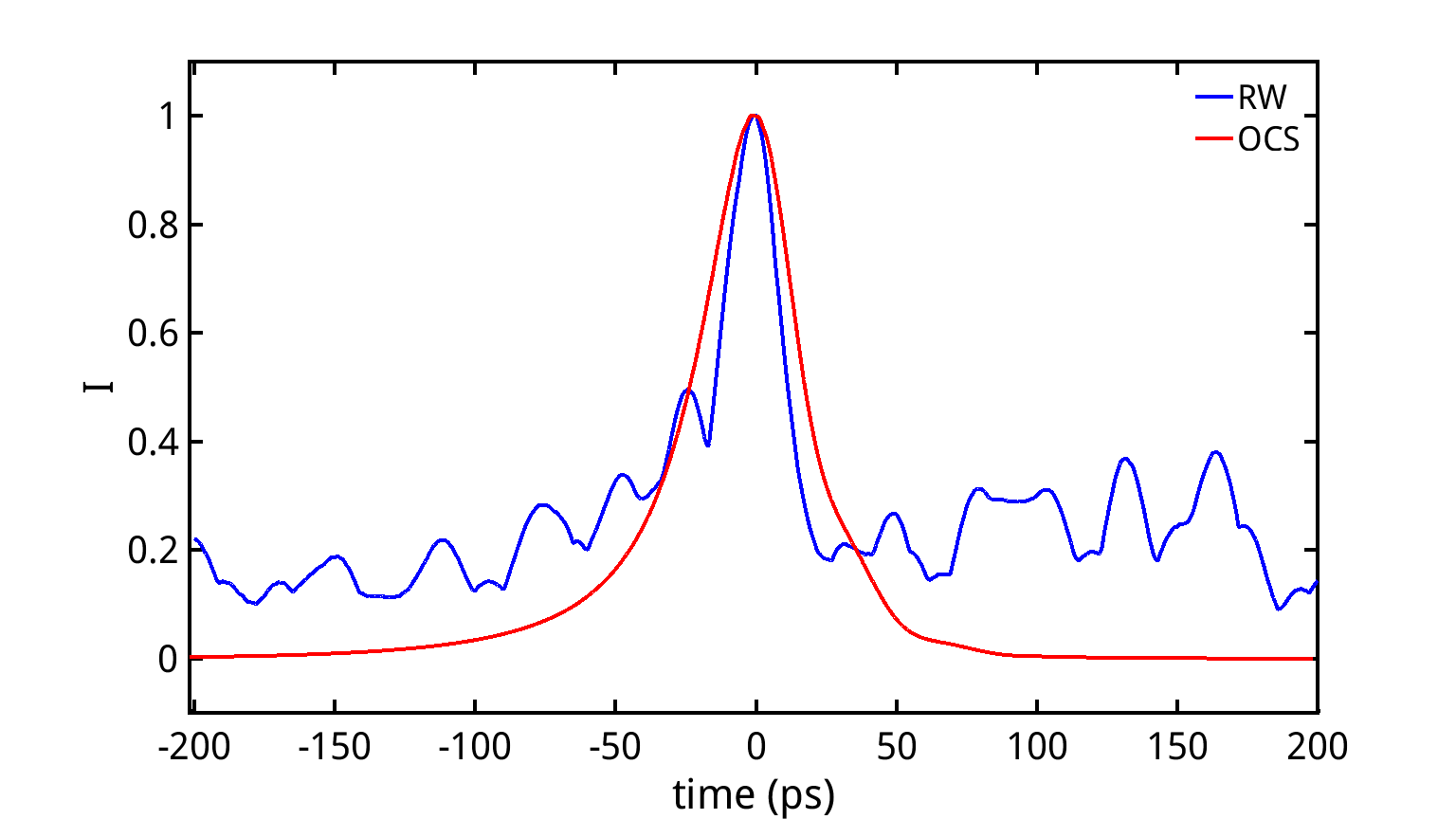}
  }
  }
  \caption{Comparison with cavity solitons. (a) Comparison of the spatial profile of a stationary cavity soliton and that of a RW, for $\mu = 5, r = 1$. Comparison of the spatial (b) and temporal (c) profiles of a self--pulsing cavity soliton and those of a RW, for $\mu = 5, r = 1.75$. The profiles are normalised to the peak intensity value.}
  \label{comparison}
\end{figure}

In Fig. \ref{comparison} (a) we present a comparison between the spatial profile of a rogue wave and a stationary cavity soliton, obtained in a parameter region where they coexist (here, $\mu = 5, r = 1$), while in Fig. \ref{comparison} (b) and (c) we compare both the spatial and temporal profiles of a RW and a self--pulsing cavity soliton, for $\mu = 5$, and $ r = 1.75$.

The very similar spatial and temporal shapes seem to indicate the same generating mechanisms for cavity solitons and RWs and that the RWs occurrence may be related to the existence of the dissipative soliton attractor in a very close parameter range.
 As for the spatial profile, the generating mechanism is related to the modulational instability of the homogeneous stationary solution, whose spatial scale is ruled by the diffraction length (depending, in turn, on the cavity length and on the wavelength of the light). Conversely, for the temporal profile, this generating mechanism is connected to the Hopf instability affecting the stationary solution (homogeneous and CS), giving rise to the well--known phenomenon of Q--switching in the plane--wave case.

\section{Conclusions}

We analysed a model for a monolithic broad-area VCSEL with an intracavity saturable absorber and introduced a new method to define and statistically analyse the \textquotedblleft events\textquotedblright, that is, the spatio--temporal maxima occurring in the transverse profile of the field intensity.

We have shown numerically the existence of rogue waves in this system according to different possible definitions and analysing different RW indicators, and we showed the best parameter choice to observe them. Furthermore, from a study of the temporal and spatial profiles, we have determined the typical temporal and spatial size (FWHM) expected for such extreme events.\\

As suggested in \cite{bonazzola} for a similar system, we believe that two-dimensional spatial effects play a crucial role in the formation of extreme events.

The same kind of analysis can be applied to different optical systems such as spatially extended, injected semiconductor lasers, such as coherently injected broad-area VCSELs \cite{oppoPRA2010,gradients} and macroscopic semiconductor ring lasers with coherent injection \cite{gustavePRL,gustavePRA}, and these works are in progress.

A future aim of the work presented here is the investigation of the predictability of rogue waves both in time and space, for example by checking if the shape of the field intensity versus space or time presents some regularities approaching a rogue wave, a development that would be very interesting especially in the framework of the hydrodynamical analogy. The identification of some typical temporal or spatial shape as a precursor of the rogue wave would allow to predict it and reduce the possible damages caused to ships or coasts.

\bibliography{References}

\end{document}